\documentclass[twocolumn, prb, showpacs]{revtex4-1}

\usepackage{amsmath, amssymb, bm, physics}
\usepackage{mathrsfs}
\usepackage{xcolor, graphicx}
\usepackage{dcolumn}
\usepackage{hyperref}
\usepackage[utf8]{inputenc} % Use UTF-8 if possible
\usepackage{comment}
\usepackage{svg}
\usepackage{float}
\usepackage{subcaption}
\usepackage{xr}
\DeclareUnicodeCharacter{0308}{\"}
\newcommand{\angstrom}{\textup{\AA}}
\begin{document}
%\captionsetup{justification=raggedright,singlelinecheck=false}
%\title{Intrinsic Magnetism Due to Charge and Spin Density Waves in 1T-CrTe$_2$ Monolayers}
%\title{Magnetic Stability and Fermi Surface Topology in Monolayer 1T-CrTe$_2$}
\title{Magnetic Stability, Fermi Surface Topology, and Spin-Correlated Dielectric Response in Monolayer 1T-CrTe$_2$}
\author{Ahmed Elrashidy}
\email{aalras2@students.towson.edu}
\affiliation{Department of Physics, Astronomy, and Geosciences, Towson University, 8000 York Road, Towson, MD 21252, USA}

\author{Jia-An Yan}
\email{jyan@towson.edu}
\affiliation{Department of Physics, Astronomy, and Geosciences, Towson University, 8000 York Road, Towson, MD 21252, USA}

\begin{abstract}
 We have carried out density-functional theory (DFT) calculations to study the magnetic stability of both ferromagnetic (FM) and anti-ferromagnetic (AFM) states in monolayer 1T-CrTe$_2$. Our results show that the AFM order is lower in energy and thus is the ground state. By tuning the lattice parameters, the AFM order can transition to the FM order, in good agreement with experimental observation. We observe a commensurate SDW alongside the previously predicted CDW, and attribute the AFM order to the SDW. This results in distinct hole and electron Fermi pockets and a pronounced optical anisotropy, suggesting quasi-one-dimensional behavior in this material.

\end{abstract}

\maketitle

%\tableofcontents

\section{Introduction\label{intro}}
%Figure 1 -----------------------------------------------------------------------------------
\begin{figure*}[tbp]
\centering
\includegraphics[width=1.00\textwidth]{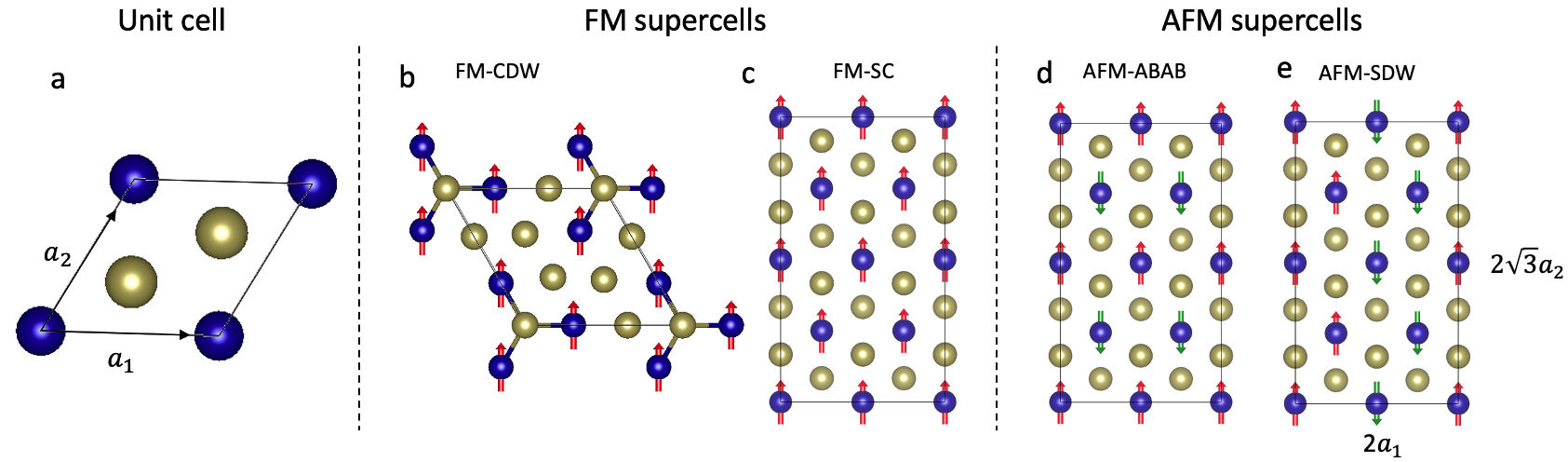}
\caption{ The unit cell of 1T-CrTe$_2$ along with the supercells used in the DFT calculations. The Cr atoms are shown in blue while the Te atoms are shown in gold. a) The unit cell and the basis vectors $a_1$ and $a_2$ are shown. b) The supercell of the CDW phase (FM-CDW). c) The supercell of the FM phase (FM-SC). d) The supercell of the abab AFM order (AFM-ABAB). e) The supercell of the SDW phase (AFM-SDW).} 
\label{fig1}
\end{figure*}

The discovery of long-range magnetic order in two-dimensional (2D) CrI$_3$ down to the monolayer threshold in 2017 has ignited a surge of interest in exploring the magnetic characteristics of 2D materials.\cite{huang2017layer} This pioneering discovery, coupled with subsequent observations of 2D magnetism in various materials, has paved the way for a plethora of potential applications in the realm of spintronics. These applications are particularly enticing due to the potential of 2D materials to serve as energy-efficient alternatives to traditional electronic devices.\cite{lin2019two}. Some spintronics applications, to name a few, include spin valves and spin field-effect transistors. \cite{khan2020recent} Additionally, 2D magnets have recently found applications in developing neuromorphic computing architectures. \cite{kwon2022memristive,zhou2021prospect}

To truly revolutionize next-generation spintronics with 2D magnetism, we need to find magnets. These magnets should have critical temperatures that are robust enough to withstand ambient conditions.. A challenge lies in the fact that many experimentally synthesized 2D magnets exhibit critical temperatures significantly below room temperature, both in their ferromagnetic (FM) and antiferromagnetic (AFM) states. This has been a significant roadblock in realizing the full potential of these materials in practical applications. For instance, the FM magnet CrI$_3$ boasts a Curie temperature ($T_C$) of 45K, while Cr$_2$Ge$_2$Te$6$ has a $T_C$ of approximately 66K.\cite{gong2017discovery,huang2017layer} On the higher end, FM order has been reported at room temperature in MnSe$_\text{x}$ films but the various $T_C$ experimental values have not been determined. \cite{o2018room} Nonetheless, the $T_C$ of MnSe$_2$ was theoretically estimated at 225K which can be further increased by applying strain to reach 330K. \cite{kan2014ferromagnetism}  

On the other hand, AFM phases have been reported in the 2D phosphorus trisulfide magnets NiPS$_3$, FePS$_3$, and MnPS$_3$ with reported N\'eel temperatures (T$_N$) of 150K, 118K, and 78K, respectively. \cite{sun2019probing,lee2016ising,kang2020coherent} More importantly, quasi-dimensional behavior was reported in the trisulfide magnets due to significant thermal and optical anisotropies \cite{kargar2020phonon,zhang2021observation,hwangbo2021highly,kim2023anisotropic}. On top of that, the Van der Walls antiferromagnet CrSBr(T$_N\approx$132K) has been shown to be quasi-one-dimensional through anisotropies in effective mass and dielectric screening. \cite{klein2023bulk} These recent experimental observations suggest that the 2D transition metal antiferromagnets would be an excellent platform for studying spin-correlated quantum phenomena in low-dimensional materials down to the 1D limit.

1T-CrTe$_2$, which in its non-magnetic phase, crystallizes in the trigonal omega-structured $\bar{\text{P}}$3m1 space group, is an ideal candidate for spintronics applications. This is due to the fact that FM and AFM phases persisting up to room temperatures have been reported in 1T-CrTe$_2$. \cite{sun2020room,zhang2021room} In the ferromagnetic state, CrTe$_2$ has been shown to have the lattice parameters $a_1$ = $a_2$ = 3.81  \AA.\cite{zhang2021room} while in the AFM order, CrTe$_2$ has been shown to have lattice parameters of $a_1$ = 3.7 \AA~ and $a_2$ = 3.4 \AA. \cite{xian2022spin} This indicates that the lattice parameters and the magnetic lattice symmetry play a significant role in the realized magnetic phase.

Many theoretical calculations based on density functional theory (DFT) have been dedicated to studying the magnetic order in CrTe$_2$. A switch between the AFM and FM phases in CrTe$_2$ monolayers due to strain has been predicted. \cite{lv2015strain,zhou2022structure} Another DFT study has concluded that the monolayer and multilayered CrTe$_2$ up to 6 layers prefer an AFM ground state and that an AFM to FM transition occurs as the number of layers increases.\cite{gao2021thickness} The emergence of a charge density wave (CDW) state has also been theoretically predicted \cite {otero2020controlled} which is not surprising as a CDW state has been experimentally observed in other metal dichalcogenides.\cite{wilson1975charge,wilson1974charge,rossnagel2011origin} Furthermore, compounds that develop a CDW state can develop a spin density wave (SDW) state depending on the total spin value.\cite{gruner1994density,rossnagel2011origin} Interestingly, co-existing CDW and SDW phases have been experimentally observed in bulk Chromium. \cite{hu2022real} 

Herein, we carried out DFT calculations to understand the contrasting magnetic behavior of 1T-CrTe$_2$ at the monolayer limit. Our calculations are not only motivated by understanding the peculiar FM and AFM phases of 1T-CrTe$_2$ but also the implications of the onset of long-range magnetic order on the dynamical, electronic, and optical properties. Hence, we start by constructing supercells corresponding to multiple suggested FM and AFM magnetic ordering in the literature and eliminate dynamically unstable supercells. We then vary the lattice parameters of dynamically stable FM and AFM phases and compare them in terms of energetic favorability to obtain a phase diagram of the possible magnetic states. A major discovery in our research is that stable FM orders are associated with CDW phases, while stable AFM orders correspond to SDW phases. Moreover, we found out that the Fermi surface corresponding to each stable phase evolves in a systematic manner. Finally, we show that Fermi nesting features are responsible for stabilizing the AFM phase and lead to a highly anisotropic optical response suggesting quasi-one-dimensional behavior. 

\section{Computational Methods \label{method}}

We performed DFT calculations using the projected augmented wave (PAW) method as implemented in the Vienna ab initio Simulation Package (VASP).\cite{kresse1996efficient,kresse1999ultrasoft} In our calculations, we adopted the Perdew-Burke-Ernzerhof (PBE) \cite{perdew1996generalized} flavor for the generalized-gradient exchange-correlation functional (GGA). 

The Brillouin zone was sampled using a 9$\times$5$\times$1 $k$-point grid mesh \cite{monkhorst1976special} for the rectangular cells and a 9$\times$9$\times$1 mesh was used for the parallelogram-shaped cell. Additionally, A 550 eV plane wave cutoff energy was used. A vacuum layer of more than 18 $\angstrom$ was applied along the $z$ direction to minimize the interactions between images of layers. Each structure was relaxed till the Hellman-Feynman forces on each atom were less than $2\times10^{-3}$ eV/$\angstrom$ and the energy convergence criterion was set to $10^{-8}$ eV.

%In these calculations, we have not considered Spin-Orbit Coupling (SOC) effects because the Magnetic Anisotropy Energy (MAE) is not being considered especially since the preferred magnetization direction of the monolayer has been established to be out of plane. \cite{liu2022structural,ji2023dimension}

In the unit cell, both the lattice parameters are equal in magnitude so that $a_1 = a_2$ as shown in Fig.~\ref{fig1}. Previous DFT calculations have shown that this ferromagnetic unit cell is not dynamically stable and exhibits imaginary phonon modes \cite{gao2021thickness,liu2022structural} unless a Hubbard $U$ parameter is used.\cite{liu2022structural} In our calculation, two FM states and two AFM states were considered using four different supercells, namely FM-SC, FM-CDW, AFM-ABAB, and AFM-SDW. The supercells have a rectangular shape and the size of each cell is $2 a_1 \times 2 \sqrt{3} a_2$ except for the parallelogram-shaped FM-CDW supercell having a size of $\sqrt{3} a_1 \times \sqrt{3} a_2$. The supercells and the orientations of the magnetic moments of the Cr atoms are shown in Fig.~\ref{fig1}. 

Naturally, the FM supercells have all magnetic moments oriented in the same direction. The magnetic moments' orientations in Fig.~\ref{fig1}(e) are typically referred to as AFM-Zigzag since the magnetic moments of neighboring Cr chains parallel to the $y$-direction form a zigzag pattern. The magnetic moments' orientations alternate in the Cr atom chains parallel to the $x$-direction in Fig.~\ref{fig1}(d) and hence the name AFM-ABAB.

For the computation of electronic band structure and density of states, we employed the Local Modified Becke-Johnson (LMBJ) meta-GGA functional. Default parameter values for $\alpha$, $\beta$, $e$, $\sigma$, and $r_{s}^{th}$, as implemented in VASP, were utilized. \cite{rauch2020local} The choice of the LMBJ functional is due to its efficacy in accurately calculating band gaps of 2D materials. This reliable performance minimizes the likelihood of incorrect metallic predictions while maintaining a reasonable computational cost. \cite{tran2021bandgap,rauch2020accurate}
\begin{table*}[tbp]
\centering
 \caption{The calculated lattice parameters, total energy E$_t$, and the energy compared to the ground state energy for AFM and FM states, respectively.} \label{tab2}
\begin{ruledtabular}
\begin{tabular}{cccccc}
  & \multicolumn{2}{c}{Lattice Parameters}  & \multicolumn{2}{c}{Magnetic Stability} & 
  \multicolumn{1}{c}{Magnetic Moment}\\
    \cline{2-3}    \cline{4-5} \cline{6-6}
   & a$_1$ (\AA) & a$_2$ (\AA) & E$_t$(eV/formula) & E-E$_{ground}$ (meV/formula) & $\mu_{B}$/Cr\\
   \cline{2-3}    \cline{4-5} \cline{6-6}
FM-SC	& 3.70	&	3.70  &	-16.26	& 20  & 2.75\\
FM-CDW	& 3.70	&   3.70  &	-16.28	& 0  & 2.72\\
 \hline
AFM-ABAB 	& 3.70	&	3.54  &	-16.30	& 40  & 2.67\\
AFM-SDW 	& 3.59	&	3.60      &	-16.34	& 0  & 2.64\\
\end{tabular}
\end{ruledtabular}
\label{table1}
\end{table*}
\begin{figure*}[tbp]
\centering
\includegraphics[width=1.00\textwidth]{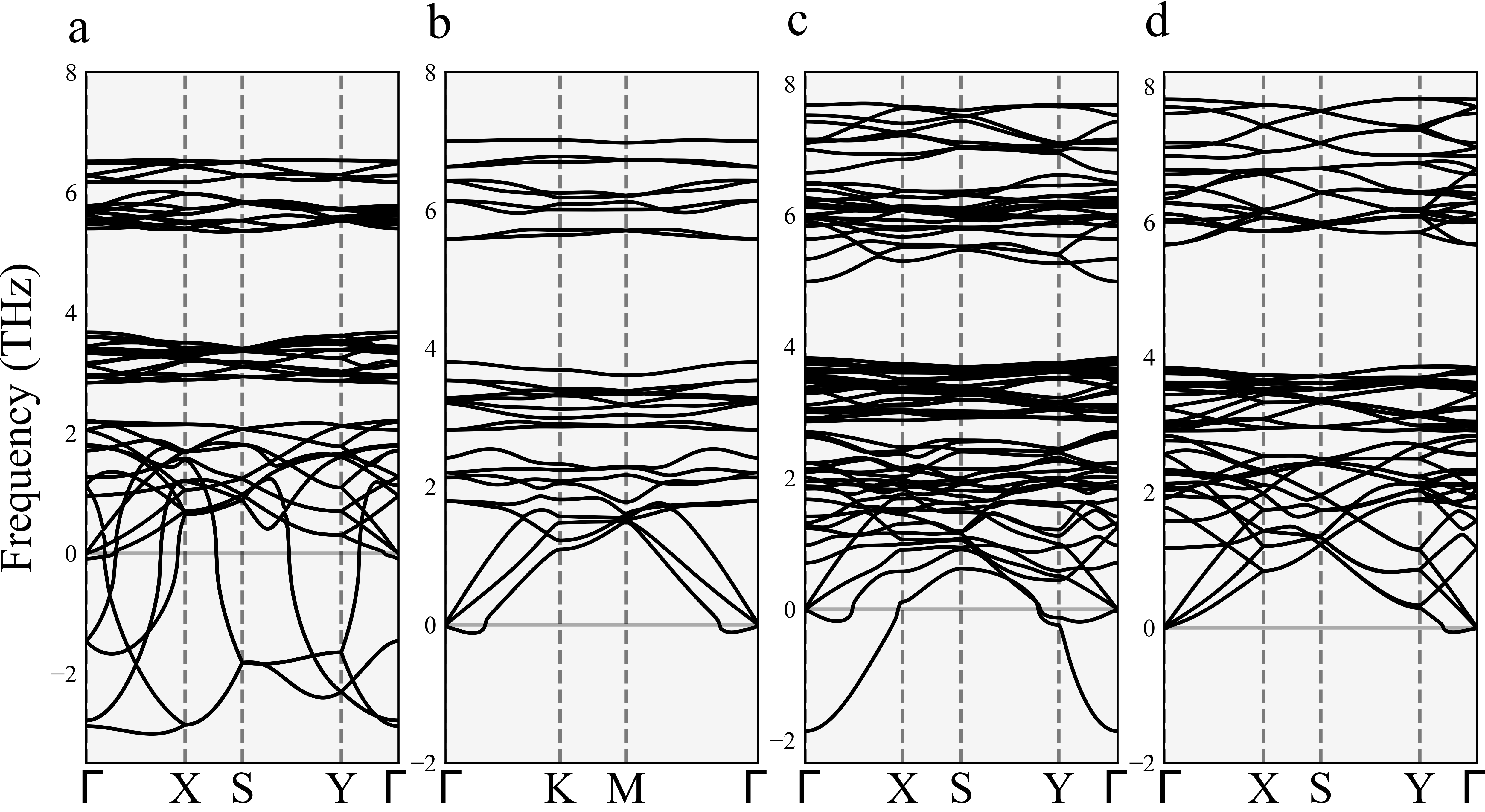}
\caption{The phonon band dispersions of the considered FM and AFM states. (a) FM-SC, (b) FM-CDW, (c) AFM-ABAB, and (d) AFM-SDW.}
\label{fig2}
\end{figure*}
\section{Results and Discussions}\label{result}
\subsection{Magnetic $\&$ Dynamical Stability}

Upon optimizing the lattice structures, we attained the optimized parameters for the four different supercells, summarized in Table~\ref{table1} alongside total energies and magnetic moments per Cr atom.

Examining the FM structures, which have consistent lattice parameters of $a_1$ = $a_2$ = 3.7 \AA, it was found that the FM-CDW state is a 20 meV lower in energy per formula than the FM-SC state. Furthermore, it is dynamically stable against FM-SC, as evidenced by the absence of imaginary phonon modes in its phonon band dispersions (Fig.~\ref{fig2}). This aligns with previous literature which reported that the CDW is the ferromagnetic ground state in monolayer CrTe$_2$. \cite{otero2020controlled}

The AFM states, however, show differing lattice parameters. The AFM-ABAB supercell displays $a_1$ = 3.7 \AA~ and $a_2$ = 3.54 \AA, while the AFM-SDW showcases $a_1$ = 3.59 \AA~ and $a_2$ = 3.6 \AA. The AFM-SDW state, with 40 meV lower in energy per formula than the AFM-ABAB state, emerges as the energetically most favorable. This state also displays dynamic stability, unlike the AFM-ABAB supercell which shows imaginary phonon modes. Therefore, the AFM-SDW state is the most stable AFM ground state of CrTe$_{2}$.

Based on our calculations, we concluded that the AFM-SDW state is the magnetic ground state for the monolayer among the FM and AFM configurations considered. This result is in agreement with the recent observation of Zigzag AFM order in the monolayer CrTe$_2$. \cite{xian2022spin} Nonetheless, several studies also reported FM ordering in bulk and low-dimensional CrTe$_2$ with lattice parameters that are larger than the reported lattice parameters of the AFM state. \cite{zhang2021room,sun2020room,freitas2015ferromagnetism,sun2021ferromagnetism,meng2021anomalous}

\begin{figure}[H]
\centering
\includegraphics[width=8cm,
  height=8cm]{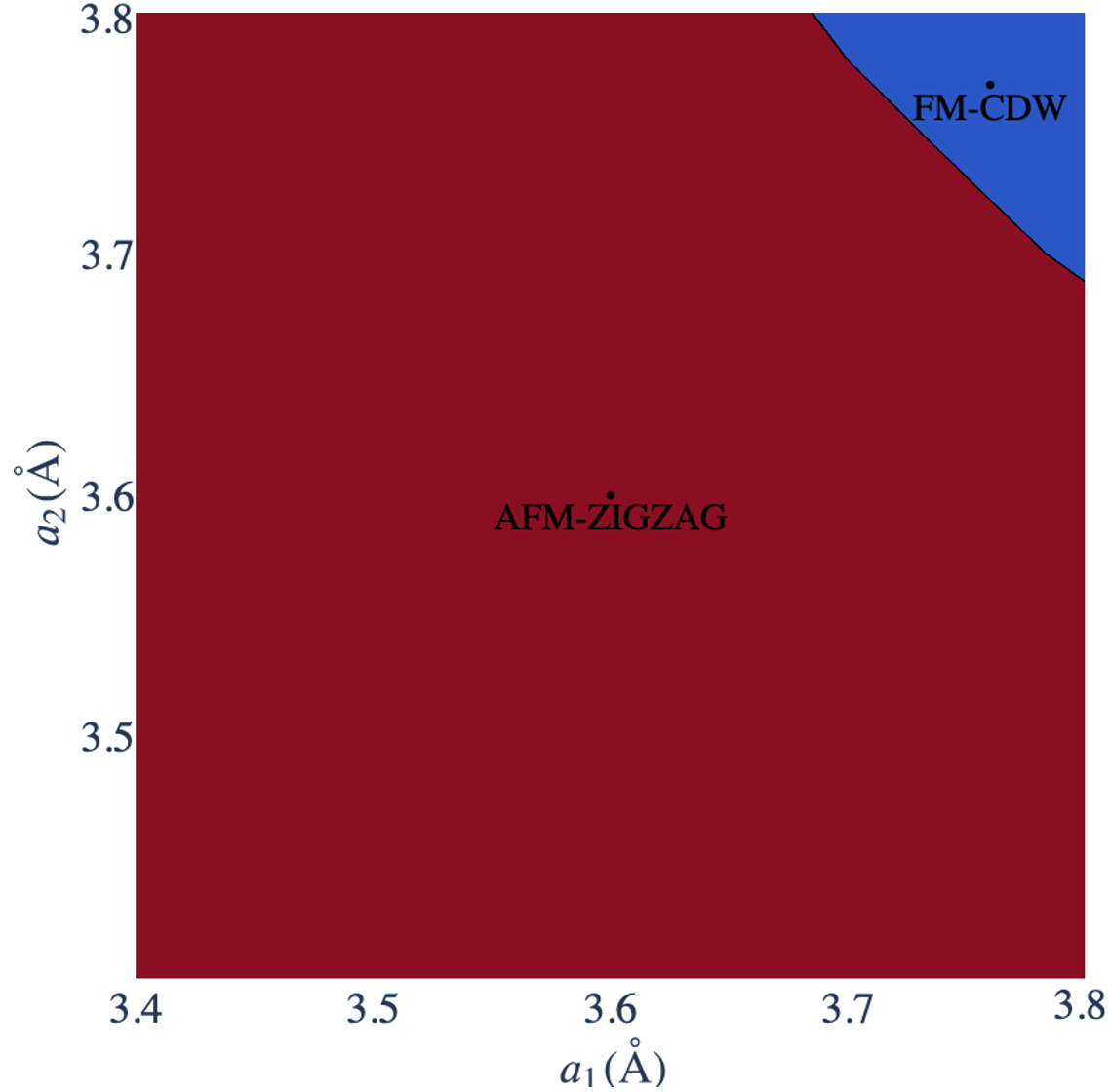} % Just stack two include graphics
 \caption{The phase diagram of the switch between AFM to FM with respect to lattice constants $a_1$ and $a_2$. The ticks on the $y$-axis and $x$-axis show the values for which the energy difference was calculated. The rest of the values were interpolated to produce the phase diagram.}
 \label{fig3}
 \end{figure}

Many DFT studies have explored how changes in lattice parameters can lead to a switch between FM and AFM states in various 2D magnetic materials. \cite{webster2018strain,hu2020enhanced,zhou2022structure, gao2021thickness}. To further clarify the magnetic order of monolayer CrTe$_2$ and whether it is possible to tune its magnetic order, we perform self-consistent calculations with different lattice parameters varying between 3.4 \AA~ and 3.8 \AA~ using the FM-CDW and AFM-SDW supercells. The obtained energies with respect to the length of lattice vectors $a_1$ and $a_2$ are compared and a phase diagram of the magnetic transitions is obtained. Fig.~\ref{fig3} shows the preferred magnetic state according to the lattice parameter values. In the red region, the AFM-SDW is more energy-efficient. In contrast, the FM-CDW is the preferred state in the blue region.  It is necessary to point out that the only considered AFM phase is due to an SDW since it allows for the emergence of AFM order in metallic systems \cite{bergeron2012breakdown,sachdev2016spin,sykora2021fluctuation,sachdev2012antiferromagnetism}. This is also supported by the fact that this AFM-SDW phase is dynamically stable.
 
This phase diagram is in excellent agreement with the reported lattice parameters of the monolayer. Specifically, in the AFM phase, the experimentally observed lattice parameters of the monolayer are $a_1$ = 3.7 \AA~ and $a_2$ = 3.4 \AA. \cite{xian2022spin} In the FM phase, the experimentally observed lattice parameters of the monolayer are $a_1$ = $a_2$ = 3.81  \AA.\cite{zhang2021room}  In the phase diagram we notice that the ferromagnetic phase tends to favor larger lattice parameters. The tendency of the ferromagnetic phase to prefer larger lattice parameters is intuitive to understand since neighboring magnetic moments align in parallel. In contrast, antiferromagnetic order, where magnetic moments can form alternating patterns, shows a tendency towards smaller lattice parameters.

\subsection{Charge \& Spin Density Waves}

Commensurate and incommensurate density waves in chromium and its alloys have been subject to multiple extensive theoretical and experimental studies. Chromium's anti-ferromagnetic behavior below its N\'eel temperature is attributed to the formation of a SDW. \cite{kubler1980spin, fawcett1994spin, fawcett1988spin,niklasson1999spin,shibatani1969spin,hirai1997magnetism,kulikov1984spin,mannix2001chromium}  More recently, a SDW in chromium has been observed in real space through spin-polarized scanning tunneling microscopy (SP-STM). \cite{hu2022real} The commensurate density waves are going to be the subject of discussion in this study since the PAW method used herein imposes periodicity and is not suitable for studying incommensurate density waves in general without special treatments. \cite{chowdhury2022computational} 

\begin{figure}[H]
\includegraphics[width=8.5 cm,
  height=6cm]{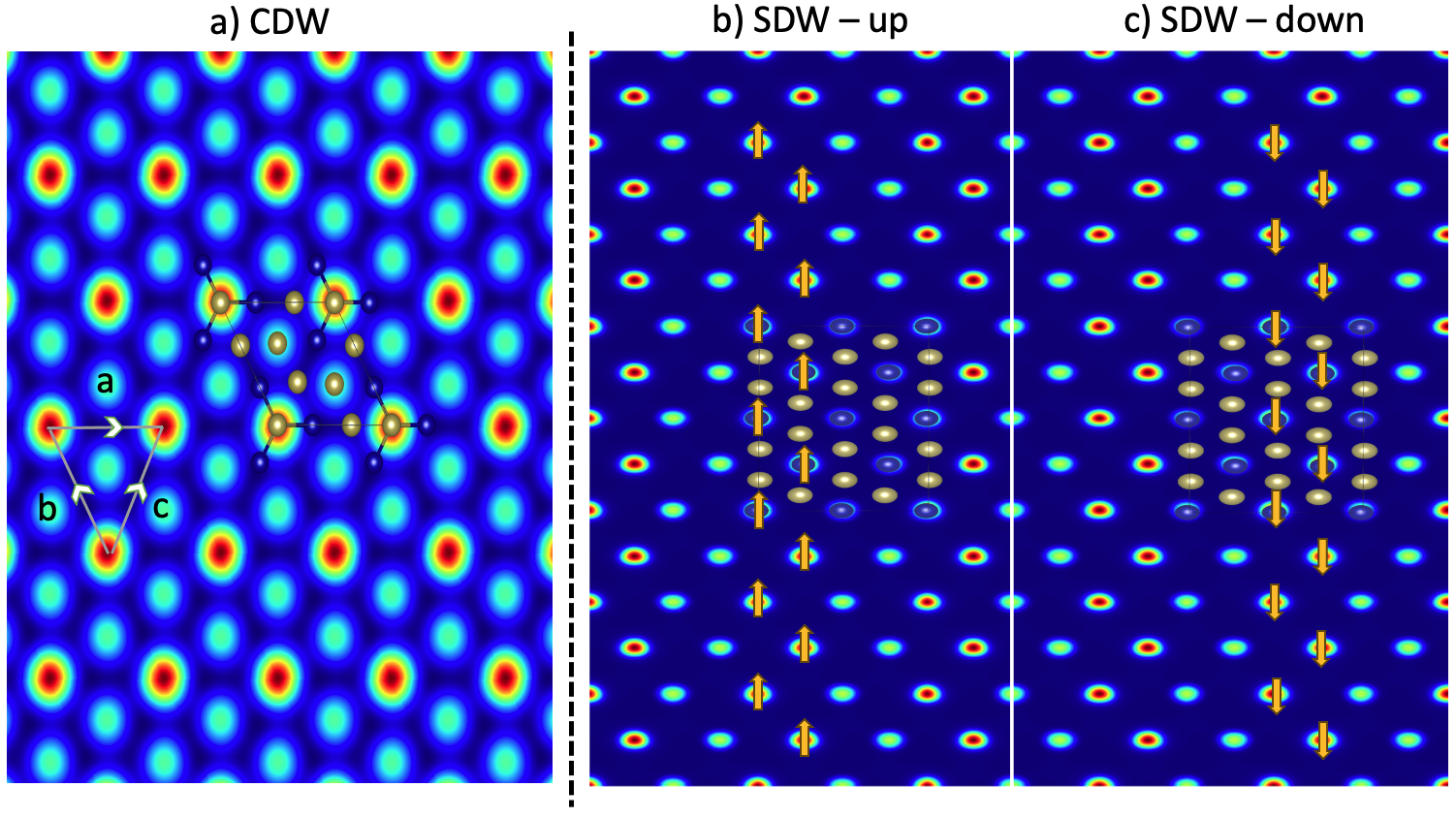} 
 \caption{Simulated STM and SP-STM images. a) STM of the incommensurate CDWs forming hexagonal concentrations. b,c) SP-STM of the spin-up (spin-down) density waves on the left (right).} \label{fig4}
\end{figure}

In non-metals, antiferromagnetism is usually addressed with spin exchange interaction models. Such models are successful at describing interacting localized magnetic moments in non-metals. However, the situation is more fluid in metals where itinerant electrons can lead to non-localized magnetic moments; when a metallic system develops a commensurate SDW, the system transitions into an AFM state \cite{blundell2003magnetism} and multiple models have been proposed to explain SDW's using both electron gas and tight-binding approximations. \cite{overhauser1962spin,gruner1994dynamics,lee1974conductivity,mizuno1959electron,huang1990imperfect}
To show the structure of the aforementioned density waves, we plot the charge density of the FM-CDW supercell and the spin densities of the AFM-SDW supercell as shown in Fig.~\ref{fig4}.  These plots simulate STM (Scanning Tunneling Microscopy) and SP-STM (Spin-Polarized Scanning Tunneling Microscopy) images of the CDW and SDW phases, respectively. The simulated images were generated using a constant height scan of 3 \AA \text{} with the exclusion of the effect of a bias voltage.

In the simulated STM image, the charge density forms hexagonal concentrations due to the overlap of three distinct incommensurate (with respect to the unit cell) CDW's. The directions of the three charge density waves are shown in Fig.~\ref{fig4}(a) by vectors $\vec{a}, \vec{b}, \vec{c}$. The three vectors are of the same length and can form an equilateral triangle. $\vec{a}$ and $\vec{b}$ are exactly the first and second lattice vectors of the supercell while $\vec{c}$ is represented by the shorter diagonal of the supercell. The incommensurate charge density waves modulate with a period of $\sqrt{3} |\vec{a_1}|$ along the directions of the three aforementioned vectors. It's worth noting that while the CDWs are incommensurate with respect to the unit cell, the supercell itself is commensurate having a $\sqrt{3} \times \sqrt{3}$ symmetry with respect to the unit cell. The switch to a CDW in phase in this material has been attributed to the activation of the 1.96 THz optical phonon mode. \cite{otero2020controlled}

The hexagonal peaks we observed have also been predicted in single-layer TMD compounds using Landau's Theory of CDW, where charge density serves as an order parameter. \cite{mcmillan1975landau} Moreover, they were also reported experimentally in 1T-TaSe$_2$ using STM \cite{chen2020strong,colonna2005mott}. 

\begin{figure}[H]
\includegraphics[width=8.55 cm,
height=10 cm]{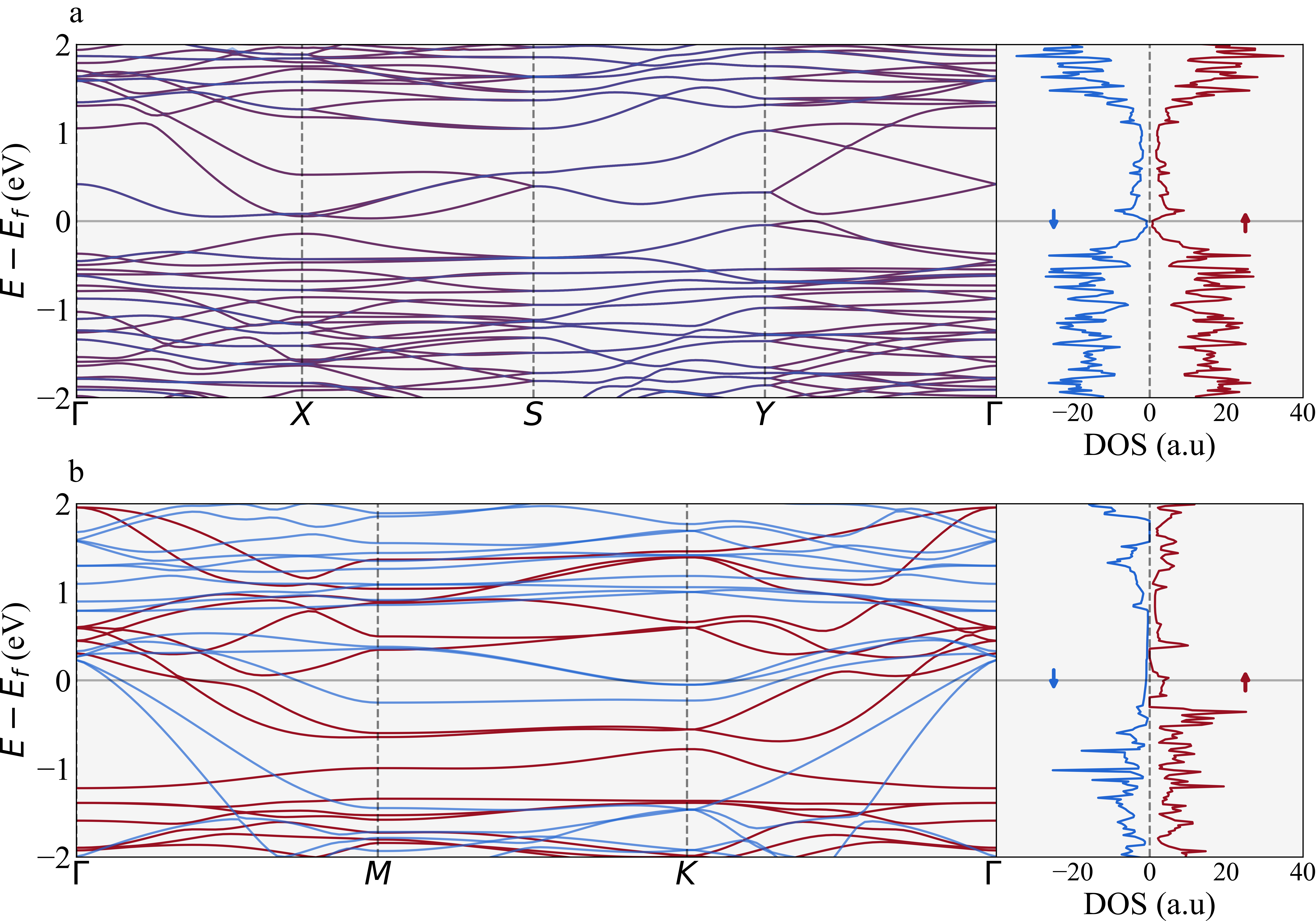}
\caption{Electronic band structures and DOS for the (a) AFM-SDW and (b) FM-CDW and phases. Spin-up (red) and down (blue) are shown.}\label{fig5}
\end{figure}

The electronic band structures and total density of states of the FM-CDW phase are shown in Fig.~\ref{fig5}. Consistent with the development of CDW, we notice an appreciable decrease in the density of states above the Fermi energy level compared to the band structures of the FM unit cell in Fig.~\ref{S1}. In the SDW phase, the SP-STM simulated images reveal distinct patterns for both spin channels, as shown in Fig.~\ref{fig4}. Unlike the CDW, the SDW phase features a commensurate density wave along the $x$-direction with a modulation period of $2|\vec{a_1}|$ and is responsible for the long-range AFM order. Notably, there is a noticeable decrease in the density of states near the Fermi energy, as seen in (b) of Fig.~\ref{fig5}. This is not indicative of a Mott transition, which is typically characterized by a full gap due to strong electron-electron interactions. Instead, this decrease in the density of states near the Fermi level may be associated with a Slater transition since it coincides with a magnetic ordering transition due to the onset of the SDW, a characteristic feature of Slater transitions. \cite{fazekas1999lecture}

\begin{figure*}[tbp]
\centering
\includegraphics[width=1.00\textwidth]{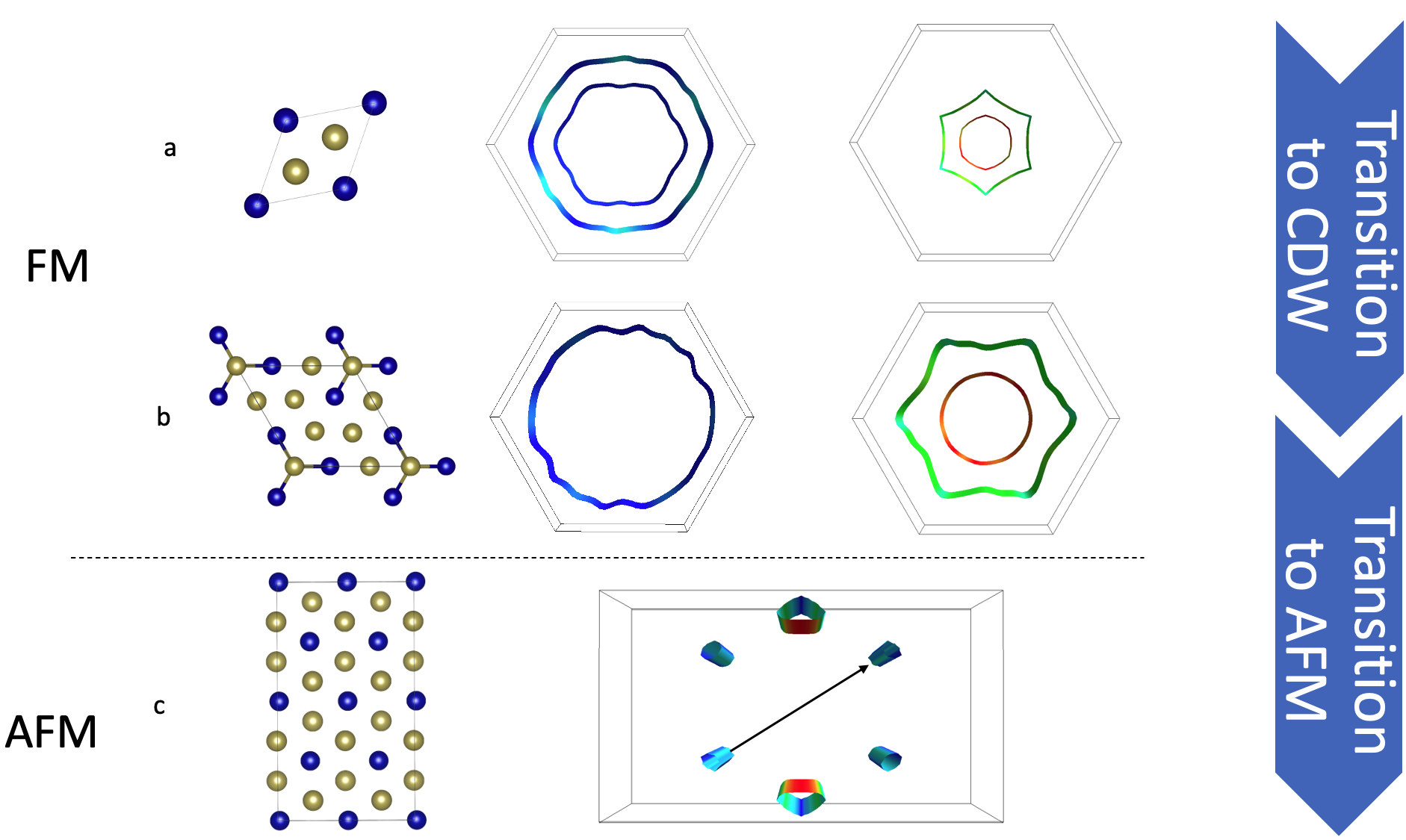}
\caption{The Fermi surfaces of the magnetic cells were generated by using FermiSurfer. \cite{kawamura2019fermisurfer} The colors indicate the Fermi velocity on an RGB scale so that the highest velocities are represented by red. (a,b) Spin-up and Spin-Down Fermi surfaces of the FM unit cell and the FM-CDW supercell, respectively. (c) Spin-degenerate Fermi surface of the AFM supercell. An example of the nesting vector connecting the hole pockets is also shown}
\label{fig6}
\end{figure*}

\subsection{Evolution of The Fermi Surface}

The Fermi surface's topology often reveals key characteristics of low-dimensional materials. This is especially true for systems with unique phases, which arise from the development of charge and spin density wave phases.\cite{bao2022spin,johannes2008fermi,johannes2006fermi,whangbo1991hidden,moore2010fermi,laverock2005fermi,knowles2020fermi,guller2016spin,sachdev2018topological,dugdale2016life}

The system's exotic density wave phases emphasize how the Fermi surface's topology plays a role in its transitions. In Fig.~\ref{fig6}, we examine the relationship between the Fermi surface and phase transitions. We start by considering the ferromagnetic unit cell and show its Fermi surface. As the transition into a CDW takes place, the Fermi surface reconstructs for both the spin-up and spin-down channels. Specifically, the Fermi surface of the spin-up electrons in the ferromagnetic CDW phase shows one less contour line indicating a reduction of the occupied electronic states at the Fermi level due to the onset of the CDW. Since this phase is also dynamically stable, it shows that the delicate interplay between electrons and phonons is ultimately necessary for stabilizing the ferromagnetic order.

The second transition highlighted in the figure is that from the FM ordering to the AFM ordering due to the onset of the SDW. As this transition manifests, the Fermi surface topology changes drastically leading to the development of Fermi pockets. This kind of Fermi surface topology is typical in metallic systems exhibiting an AFM order and the theory behind the transition has been discussed extensively in the literature. \cite{abanov2000spin,bauer2020hierarchy,sachdev2012antiferromagnetism,eberlein2016fermi,sykora2021fluctuation} As seen in Fig.~\ref{fig6} when the transition to an AFM occurs, the monolayer transitions from being a metal with a large Fermi surface to a metallic state exhibiting electron and hole pockets due to the development of the SDW order at the wavevector $\vec{k}$ = ($\pi$,$\pi$). While it is expected that the increase of the strength of the U parameter in a Hubbard-like model leads to this evolution from a larger Fermi surface to a ``pocketed" Fermi surface, \cite{sachdev2012antiferromagnetism} our calculations demonstrate that such evolution can be observed without explicitly accounting for on-site repulsions.

\begin{figure}[H]
\includegraphics[width=8.55 cm,
height= 8.55 cm]{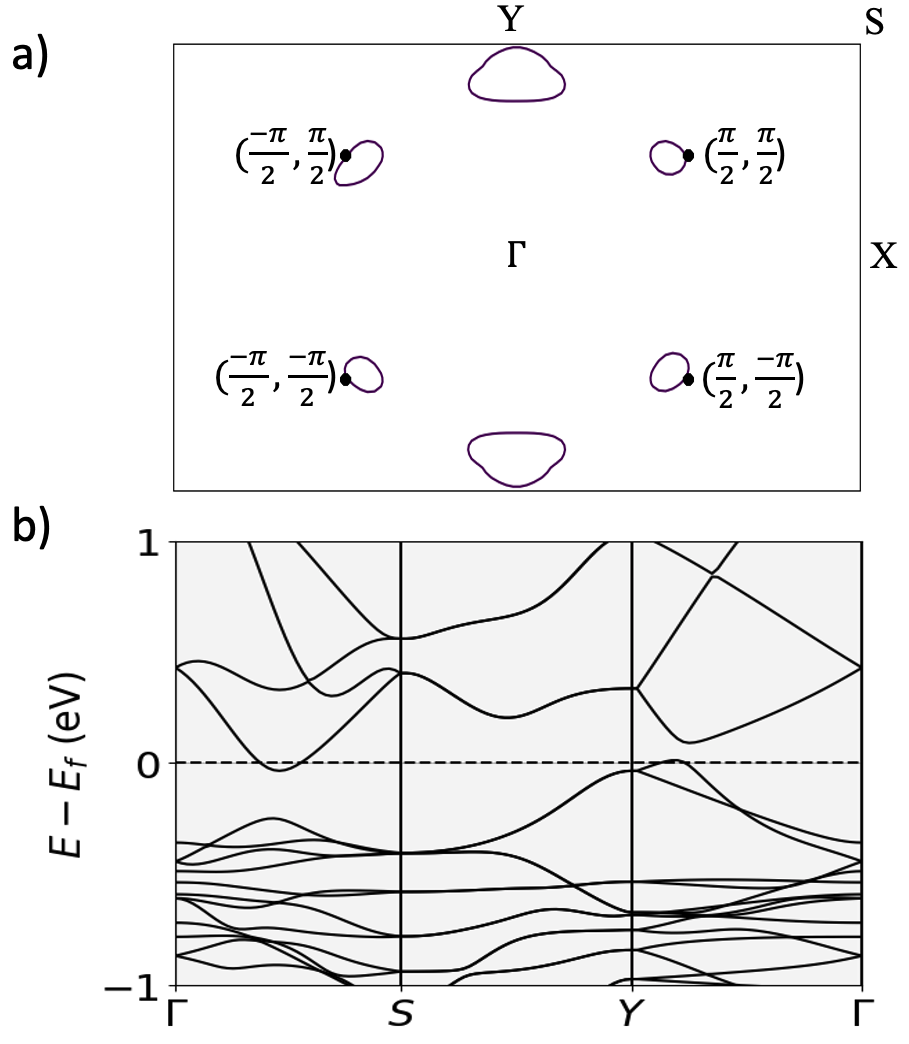}
\caption{a) The Fermi surface with hole and electron pockets on the right. b) The band structures along the path relevant to the Fermi surface of the SDW}
\label{fig7}
\end{figure}

 The hole pockets are due to contributions above the Fermi energy along the $\Gamma$-S path while the electron pockets are due to contributions along the $\Gamma$-Y path. To better depict this, we have plotted the electronic band structures along these paths in Fig.~\ref{fig7}. In the same figure, we also highlight the location of the expected nested hole pockets around ($\frac{\pi}{2}$,$\frac{\pi}{2}$) and symmetry-related points due to the nesting vector $\vec{k}$ = ($\pi$,$\pi$). \cite{abanov2000spin}

\subsection{Anisotropic Optical Response}

SDW phases typically arise in highly anisotropic metals. \cite{gruner1994density} This highly anisotropic character presents itself in the optical response of quasi-one-dimensional chains and has been shown for the organic linear-chain compound (TMTSF)$_2$PF$_6$ exhibiting a SDW phase. \cite{jacobsen1982optical} Of particular note is the large optical anisotropy resulting in linear dichroism, recently observed in zigzag antiferromagnets such as FePS$_3$ and NiPS$_3$. \cite{zhang2021observation,hwangbo2021highly,kim2023anisotropic} 

To probe the optical properties of the SDW phase we performed time-dependent DFT calculation to obtain the complex dielectric tensor $\vectorbold{\epsilon} = \vectorbold{\epsilon} ' + i \vectorbold{\epsilon}''$.

The real part of the tensor element along the $(\alpha \beta)$ direction is given by the Kramers-Kronig transformation:
\begin{equation}
    \epsilon _{{\alpha \beta }}'(\omega )=1+{\frac  {2}{\pi }}P\int _{0}^{{\infty }}{\frac  {\epsilon _{{\alpha \beta }}^{{(2)}}(\omega ')\omega '}{\omega '^{2}-\omega ^{2}+i\eta }}d\omega
\end{equation}

where $\alpha$ and $\beta$ represent the Cartesian coordinates ($x$,$y$,$z$) and $P$ is the principle value. 

The imaginary part of the tensor element is determined by summation over empty states as follows:
\begin{align}
\epsilon _{{\alpha \beta }}^{{(2)}}\left(\omega \right) & ={\frac  {4\pi ^{2}e^{2}}{\Omega }}{\mathrm  {lim}}_{{q\rightarrow 0}}{\frac  {1}{q^{2}}} \nonumber \\
& \times \sum _{{c,v,{\mathbf  {k}}}}2w_{{\mathbf  {k}}}\delta (\epsilon _{{c{\mathbf  {k}}}}-\epsilon _{{v{\mathbf  {k}}}}-\omega ) \nonumber \\
& \times \langle u_{{c{\mathbf  {k}}+{\mathbf  {e}}_{\alpha }q}}|u_{{v{\mathbf  {k}}}}\rangle \langle u_{{v{\mathbf  {k}}}}|u_{{c{\mathbf  {k}}+{\mathbf  {e}}_{\beta }q}}\rangle 
\end{align}

where $c$ refers to conduction bands and $v$ refers to valence bands while $u_{{c{\mathbf{k}}}}$ is the cell periodic part of the orbitals at the k-point.

\begin{figure}[H]
\centering
\includegraphics[width=8 cm,
height=10cm]{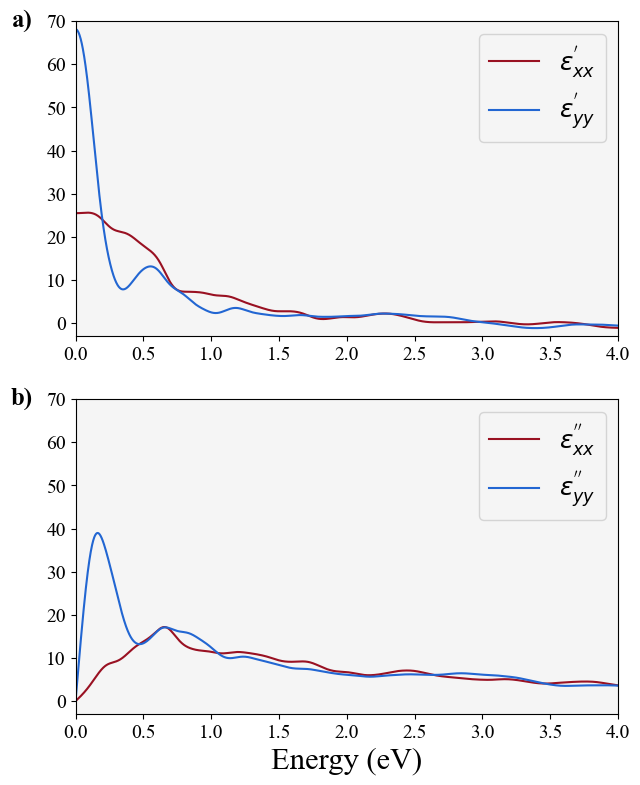}
\caption{Anisotropic optical response: The real and imaginary parts of the frequency-dependent dielectric tensor $\epsilon_{xx}$ and $\epsilon_{yy}$, reflecting anisotropic conductivity in the quasi-one-dimensional system. The real part is shown in (a) and the imaginary part is shown in (b)}
\label{fig8}
\end{figure}

We then plot the real and imaginary parts of the $\epsilon_{xx}$ and $\epsilon_{yy}$ to show the optical response along the $x$ and $y$ directions, respectively. The plotted results of these calculations, also shown in Fig.~\ref{fig8}, reveal the presence of a pronounced optical anisotropy that is not present in the ferromagnetic phases of the unit cell or the CDW supercell as seen in Fig.~\ref{S2} and Fig.~\ref{S3}. Specifically, along the $y$ direction, the real $(\epsilon_{yy}')$ and imaginary $(\epsilon_{yy}'')$ parts show large resonant responses near 0 eV. These resonances along the $y$ direction are likely due to contributions from the larger electron pockets near the Fermi surface along the $\Gamma$-Y path. The anisotropic behavior indicates that the AFM state is quasi-one-dimensional due to the SDW causing the out-phase modulation along neighboring chains leading to the Zigzag pattern of the AFM order.

% In the considered SDW-AFM phase, the modulation of the SDW along neighboring chains is out of phase by $|2\vec{a_1}|$ which is the magnitude of the lattice vector in the x direction.

\section{Conclusions}

In this work, we have explored the intricate magnetic properties of 1T-CrTe$_2$ monolayers, a potential candidate for high-temperature intrinsic magnetism. By employing density-functional theory (DFT), we have investigated the magnetic and dynamic stability of both ferromagnetic (FM) and anti-ferromagnetic (AFM) states and their relationship with lattice parameters. We have found that charge and spin density waves are responsible for stabilizing the FM and AFM magnetic orders, respectively. 

Our results suggest that the AFM state appears to be the ground state for 1T-CrTe$_2$ monolayers. We have also demonstrated that magnetic ordering can be influenced by adjusting the lattice parameters and hence allowing transitions between FM-CDW and AFM-SDW phases. This tunability of magnetism could potentially pave the way for the application of 1T-CrTe$_2$ in spintronic devices, where the manipulation of magnetic states is essential.

A key discovery of our study is the prediction of a commensurate spin density wave (SDW), which we believe is responsible for the AFM order. This SDW exhibits significant anisotropy, leading to quasi-one-dimensional behavior. Furthermore, we observed that the anisotropy of this quasi-one-dimensional SDW extends to the material's optical response. This finding presents an intriguing direction for future research and suggests that understanding the relationship between magnetic and optical properties could lead to the development of novel magnetically tunable optoelectronic devices.

\section*{Acknowledgements}
This work used EXPANSE at the San Diego Supercomputing Center (SDSC) through allocations DMR16088 and PHY220161 from the Advanced Cyberinfrastructure Coordination Ecosystem: Services \& Support (ACCESS) program. We acknowledge support from the NSF grant DMR 1709781 and support from the Fisher General Endowment and SET grants from the Jess and Mildred Fisher College of Science and Mathematics at Towson University.

\newpage
\bibliography{biblo}

\begin{thebibliography}{79}%
\makeatletter
\providecommand \@ifxundefined [1]{%
 \@ifx{#1\undefined}
}%
\providecommand \@ifnum [1]{%
 \ifnum #1\expandafter \@firstoftwo
 \else \expandafter \@secondoftwo
 \fi
}%
\providecommand \@ifx [1]{%
 \ifx #1\expandafter \@firstoftwo
 \else \expandafter \@secondoftwo
 \fi
}%
\providecommand \natexlab [1]{#1}%
\providecommand \enquote  [1]{``#1''}%
\providecommand \bibnamefont  [1]{#1}%
\providecommand \bibfnamefont [1]{#1}%
\providecommand \citenamefont [1]{#1}%
\providecommand \href@noop [0]{\@secondoftwo}%
\providecommand \href [0]{\begingroup \@sanitize@url \@href}%
\providecommand \@href[1]{\@@startlink{#1}\@@href}%
\providecommand \@@href[1]{\endgroup#1\@@endlink}%
\providecommand \@sanitize@url [0]{\catcode `\\12\catcode `\$12\catcode
  `\&12\catcode `\#12\catcode `\^12\catcode `\_12\catcode `\%12\relax}%
\providecommand \@@startlink[1]{}%
\providecommand \@@endlink[0]{}%
\providecommand \url  [0]{\begingroup\@sanitize@url \@url }%
\providecommand \@url [1]{\endgroup\@href {#1}{\urlprefix }}%
\providecommand \urlprefix  [0]{URL }%
\providecommand \Eprint [0]{\href }%
\providecommand \doibase [0]{http://dx.doi.org/}%
\providecommand \selectlanguage [0]{\@gobble}%
\providecommand \bibinfo  [0]{\@secondoftwo}%
\providecommand \bibfield  [0]{\@secondoftwo}%
\providecommand \translation [1]{[#1]}%
\providecommand \BibitemOpen [0]{}%
\providecommand \bibitemStop [0]{}%
\providecommand \bibitemNoStop [0]{.\EOS\space}%
\providecommand \EOS [0]{\spacefactor3000\relax}%
\providecommand \BibitemShut  [1]{\csname bibitem#1\endcsname}%
\let\auto@bib@innerbib\@empty
%</preamble>
\bibitem [{\citenamefont {Huang}\ \emph {et~al.}(2017)\citenamefont {Huang},
  \citenamefont {Clark}, \citenamefont {Navarro-Moratalla}, \citenamefont
  {Klein}, \citenamefont {Cheng}, \citenamefont {Seyler}, \citenamefont
  {Zhong}, \citenamefont {Schmidgall}, \citenamefont {McGuire}, \citenamefont
  {Cobden} \emph {et~al.}}]{huang2017layer}%
  \BibitemOpen
  \bibfield  {author} {\bibinfo {author} {\bibfnamefont {B.}~\bibnamefont
  {Huang}}, \bibinfo {author} {\bibfnamefont {G.}~\bibnamefont {Clark}},
  \bibinfo {author} {\bibfnamefont {E.}~\bibnamefont {Navarro-Moratalla}},
  \bibinfo {author} {\bibfnamefont {D.~R.}\ \bibnamefont {Klein}}, \bibinfo
  {author} {\bibfnamefont {R.}~\bibnamefont {Cheng}}, \bibinfo {author}
  {\bibfnamefont {K.~L.}\ \bibnamefont {Seyler}}, \bibinfo {author}
  {\bibfnamefont {D.}~\bibnamefont {Zhong}}, \bibinfo {author} {\bibfnamefont
  {E.}~\bibnamefont {Schmidgall}}, \bibinfo {author} {\bibfnamefont {M.~A.}\
  \bibnamefont {McGuire}}, \bibinfo {author} {\bibfnamefont {D.~H.}\
  \bibnamefont {Cobden}},  \emph {et~al.},\ }\href@noop {} {\bibfield
  {journal} {\bibinfo  {journal} {Nature}\ }\textbf {\bibinfo {volume} {546}},\
  \bibinfo {pages} {270} (\bibinfo {year} {2017})}\BibitemShut {NoStop}%
\bibitem [{\citenamefont {Lin}\ \emph {et~al.}(2019)\citenamefont {Lin},
  \citenamefont {Yang}, \citenamefont {Wang},\ and\ \citenamefont
  {Zhao}}]{lin2019two}%
  \BibitemOpen
  \bibfield  {author} {\bibinfo {author} {\bibfnamefont {X.}~\bibnamefont
  {Lin}}, \bibinfo {author} {\bibfnamefont {W.}~\bibnamefont {Yang}}, \bibinfo
  {author} {\bibfnamefont {K.~L.}\ \bibnamefont {Wang}}, \ and\ \bibinfo
  {author} {\bibfnamefont {W.}~\bibnamefont {Zhao}},\ }\href@noop {} {\bibfield
   {journal} {\bibinfo  {journal} {Nature Electronics}\ }\textbf {\bibinfo
  {volume} {2}},\ \bibinfo {pages} {274} (\bibinfo {year} {2019})}\BibitemShut
  {NoStop}%
\bibitem [{\citenamefont {Khan}\ \emph {et~al.}(2020)\citenamefont {Khan},
  \citenamefont {Obaidulla}, \citenamefont {Habib}, \citenamefont {Gayen},
  \citenamefont {Liang}, \citenamefont {Wang},\ and\ \citenamefont
  {Xu}}]{khan2020recent}%
  \BibitemOpen
  \bibfield  {author} {\bibinfo {author} {\bibfnamefont {Y.}~\bibnamefont
  {Khan}}, \bibinfo {author} {\bibfnamefont {S.~M.}\ \bibnamefont {Obaidulla}},
  \bibinfo {author} {\bibfnamefont {M.~R.}\ \bibnamefont {Habib}}, \bibinfo
  {author} {\bibfnamefont {A.}~\bibnamefont {Gayen}}, \bibinfo {author}
  {\bibfnamefont {T.}~\bibnamefont {Liang}}, \bibinfo {author} {\bibfnamefont
  {X.}~\bibnamefont {Wang}}, \ and\ \bibinfo {author} {\bibfnamefont
  {M.}~\bibnamefont {Xu}},\ }\href@noop {} {\bibfield  {journal} {\bibinfo
  {journal} {Nano Today}\ }\textbf {\bibinfo {volume} {34}},\ \bibinfo {pages}
  {100902} (\bibinfo {year} {2020})}\BibitemShut {NoStop}%
\bibitem [{\citenamefont {Kwon}\ \emph {et~al.}(2022)\citenamefont {Kwon},
  \citenamefont {Baek}, \citenamefont {Hong}, \citenamefont {Kim},\ and\
  \citenamefont {Jang}}]{kwon2022memristive}%
  \BibitemOpen
  \bibfield  {author} {\bibinfo {author} {\bibfnamefont {K.~C.}\ \bibnamefont
  {Kwon}}, \bibinfo {author} {\bibfnamefont {J.~H.}\ \bibnamefont {Baek}},
  \bibinfo {author} {\bibfnamefont {K.}~\bibnamefont {Hong}}, \bibinfo {author}
  {\bibfnamefont {S.~Y.}\ \bibnamefont {Kim}}, \ and\ \bibinfo {author}
  {\bibfnamefont {H.~W.}\ \bibnamefont {Jang}},\ }\href@noop {} {\bibfield
  {journal} {\bibinfo  {journal} {Nano-Micro Letters}\ }\textbf {\bibinfo
  {volume} {14}},\ \bibinfo {pages} {1} (\bibinfo {year} {2022})}\BibitemShut
  {NoStop}%
\bibitem [{\citenamefont {Zhou}\ and\ \citenamefont
  {Chen}(2021)}]{zhou2021prospect}%
  \BibitemOpen
  \bibfield  {author} {\bibinfo {author} {\bibfnamefont {J.}~\bibnamefont
  {Zhou}}\ and\ \bibinfo {author} {\bibfnamefont {J.}~\bibnamefont {Chen}},\
  }\href@noop {} {\bibfield  {journal} {\bibinfo  {journal} {Advanced
  Electronic Materials}\ }\textbf {\bibinfo {volume} {7}},\ \bibinfo {pages}
  {2100465} (\bibinfo {year} {2021})}\BibitemShut {NoStop}%
\bibitem [{\citenamefont {Gong}\ \emph {et~al.}(2017)\citenamefont {Gong},
  \citenamefont {Li}, \citenamefont {Li}, \citenamefont {Ji}, \citenamefont
  {Stern}, \citenamefont {Xia}, \citenamefont {Cao}, \citenamefont {Bao},
  \citenamefont {Wang}, \citenamefont {Wang} \emph
  {et~al.}}]{gong2017discovery}%
  \BibitemOpen
  \bibfield  {author} {\bibinfo {author} {\bibfnamefont {C.}~\bibnamefont
  {Gong}}, \bibinfo {author} {\bibfnamefont {L.}~\bibnamefont {Li}}, \bibinfo
  {author} {\bibfnamefont {Z.}~\bibnamefont {Li}}, \bibinfo {author}
  {\bibfnamefont {H.}~\bibnamefont {Ji}}, \bibinfo {author} {\bibfnamefont
  {A.}~\bibnamefont {Stern}}, \bibinfo {author} {\bibfnamefont
  {Y.}~\bibnamefont {Xia}}, \bibinfo {author} {\bibfnamefont {T.}~\bibnamefont
  {Cao}}, \bibinfo {author} {\bibfnamefont {W.}~\bibnamefont {Bao}}, \bibinfo
  {author} {\bibfnamefont {C.}~\bibnamefont {Wang}}, \bibinfo {author}
  {\bibfnamefont {Y.}~\bibnamefont {Wang}},  \emph {et~al.},\ }\href@noop {}
  {\bibfield  {journal} {\bibinfo  {journal} {Nature}\ }\textbf {\bibinfo
  {volume} {546}},\ \bibinfo {pages} {265} (\bibinfo {year}
  {2017})}\BibitemShut {NoStop}%
\bibitem [{\citenamefont {O’Hara}\ \emph {et~al.}(2018)\citenamefont
  {O’Hara}, \citenamefont {Zhu}, \citenamefont {Trout}, \citenamefont
  {Ahmed}, \citenamefont {Luo}, \citenamefont {Lee}, \citenamefont {Brenner},
  \citenamefont {Rajan}, \citenamefont {Gupta}, \citenamefont {McComb} \emph
  {et~al.}}]{o2018room}%
  \BibitemOpen
  \bibfield  {author} {\bibinfo {author} {\bibfnamefont {D.~J.}\ \bibnamefont
  {O’Hara}}, \bibinfo {author} {\bibfnamefont {T.}~\bibnamefont {Zhu}},
  \bibinfo {author} {\bibfnamefont {A.~H.}\ \bibnamefont {Trout}}, \bibinfo
  {author} {\bibfnamefont {A.~S.}\ \bibnamefont {Ahmed}}, \bibinfo {author}
  {\bibfnamefont {Y.~K.}\ \bibnamefont {Luo}}, \bibinfo {author} {\bibfnamefont
  {C.~H.}\ \bibnamefont {Lee}}, \bibinfo {author} {\bibfnamefont {M.~R.}\
  \bibnamefont {Brenner}}, \bibinfo {author} {\bibfnamefont {S.}~\bibnamefont
  {Rajan}}, \bibinfo {author} {\bibfnamefont {J.~A.}\ \bibnamefont {Gupta}},
  \bibinfo {author} {\bibfnamefont {D.~W.}\ \bibnamefont {McComb}},  \emph
  {et~al.},\ }\href@noop {} {\bibfield  {journal} {\bibinfo  {journal} {Nano
  letters}\ }\textbf {\bibinfo {volume} {18}},\ \bibinfo {pages} {3125}
  (\bibinfo {year} {2018})}\BibitemShut {NoStop}%
\bibitem [{\citenamefont {Kan}\ \emph {et~al.}(2014)\citenamefont {Kan},
  \citenamefont {Adhikari},\ and\ \citenamefont {Sun}}]{kan2014ferromagnetism}%
  \BibitemOpen
  \bibfield  {author} {\bibinfo {author} {\bibfnamefont {M.}~\bibnamefont
  {Kan}}, \bibinfo {author} {\bibfnamefont {S.}~\bibnamefont {Adhikari}}, \
  and\ \bibinfo {author} {\bibfnamefont {Q.}~\bibnamefont {Sun}},\ }\href@noop
  {} {\bibfield  {journal} {\bibinfo  {journal} {Physical Chemistry Chemical
  Physics}\ }\textbf {\bibinfo {volume} {16}},\ \bibinfo {pages} {4990}
  (\bibinfo {year} {2014})}\BibitemShut {NoStop}%
\bibitem [{\citenamefont {Sun}\ \emph {et~al.}(2019)\citenamefont {Sun},
  \citenamefont {Tan}, \citenamefont {Liu}, \citenamefont {Gao},\ and\
  \citenamefont {Zhang}}]{sun2019probing}%
  \BibitemOpen
  \bibfield  {author} {\bibinfo {author} {\bibfnamefont {Y.-J.}\ \bibnamefont
  {Sun}}, \bibinfo {author} {\bibfnamefont {Q.-H.}\ \bibnamefont {Tan}},
  \bibinfo {author} {\bibfnamefont {X.-L.}\ \bibnamefont {Liu}}, \bibinfo
  {author} {\bibfnamefont {Y.-F.}\ \bibnamefont {Gao}}, \ and\ \bibinfo
  {author} {\bibfnamefont {J.}~\bibnamefont {Zhang}},\ }\href@noop {}
  {\bibfield  {journal} {\bibinfo  {journal} {The Journal of Physical Chemistry
  Letters}\ }\textbf {\bibinfo {volume} {10}},\ \bibinfo {pages} {3087}
  (\bibinfo {year} {2019})}\BibitemShut {NoStop}%
\bibitem [{\citenamefont {Lee}\ \emph {et~al.}(2016)\citenamefont {Lee},
  \citenamefont {Lee}, \citenamefont {Ryoo}, \citenamefont {Kang},
  \citenamefont {Kim}, \citenamefont {Kim}, \citenamefont {Park}, \citenamefont
  {Park},\ and\ \citenamefont {Cheong}}]{lee2016ising}%
  \BibitemOpen
  \bibfield  {author} {\bibinfo {author} {\bibfnamefont {J.-U.}\ \bibnamefont
  {Lee}}, \bibinfo {author} {\bibfnamefont {S.}~\bibnamefont {Lee}}, \bibinfo
  {author} {\bibfnamefont {J.~H.}\ \bibnamefont {Ryoo}}, \bibinfo {author}
  {\bibfnamefont {S.}~\bibnamefont {Kang}}, \bibinfo {author} {\bibfnamefont
  {T.~Y.}\ \bibnamefont {Kim}}, \bibinfo {author} {\bibfnamefont
  {P.}~\bibnamefont {Kim}}, \bibinfo {author} {\bibfnamefont {C.-H.}\
  \bibnamefont {Park}}, \bibinfo {author} {\bibfnamefont {J.-G.}\ \bibnamefont
  {Park}}, \ and\ \bibinfo {author} {\bibfnamefont {H.}~\bibnamefont
  {Cheong}},\ }\href@noop {} {\bibfield  {journal} {\bibinfo  {journal} {Nano
  letters}\ }\textbf {\bibinfo {volume} {16}},\ \bibinfo {pages} {7433}
  (\bibinfo {year} {2016})}\BibitemShut {NoStop}%
\bibitem [{\citenamefont {Kang}\ \emph {et~al.}(2020)\citenamefont {Kang},
  \citenamefont {Kim}, \citenamefont {Kim}, \citenamefont {Kim}, \citenamefont
  {Sim}, \citenamefont {Lee}, \citenamefont {Lee}, \citenamefont {Park},
  \citenamefont {Yun}, \citenamefont {Kim} \emph {et~al.}}]{kang2020coherent}%
  \BibitemOpen
  \bibfield  {author} {\bibinfo {author} {\bibfnamefont {S.}~\bibnamefont
  {Kang}}, \bibinfo {author} {\bibfnamefont {K.}~\bibnamefont {Kim}}, \bibinfo
  {author} {\bibfnamefont {B.~H.}\ \bibnamefont {Kim}}, \bibinfo {author}
  {\bibfnamefont {J.}~\bibnamefont {Kim}}, \bibinfo {author} {\bibfnamefont
  {K.~I.}\ \bibnamefont {Sim}}, \bibinfo {author} {\bibfnamefont {J.-U.}\
  \bibnamefont {Lee}}, \bibinfo {author} {\bibfnamefont {S.}~\bibnamefont
  {Lee}}, \bibinfo {author} {\bibfnamefont {K.}~\bibnamefont {Park}}, \bibinfo
  {author} {\bibfnamefont {S.}~\bibnamefont {Yun}}, \bibinfo {author}
  {\bibfnamefont {T.}~\bibnamefont {Kim}},  \emph {et~al.},\ }\href@noop {}
  {\bibfield  {journal} {\bibinfo  {journal} {Nature}\ }\textbf {\bibinfo
  {volume} {583}},\ \bibinfo {pages} {785} (\bibinfo {year}
  {2020})}\BibitemShut {NoStop}%
\bibitem [{\citenamefont {Kargar}\ \emph {et~al.}(2020)\citenamefont {Kargar},
  \citenamefont {Coleman}, \citenamefont {Ghosh}, \citenamefont {Lee},
  \citenamefont {Gomez}, \citenamefont {Liu}, \citenamefont {Magana},
  \citenamefont {Barani}, \citenamefont {Mohammadzadeh}, \citenamefont
  {Debnath} \emph {et~al.}}]{kargar2020phonon}%
  \BibitemOpen
  \bibfield  {author} {\bibinfo {author} {\bibfnamefont {F.}~\bibnamefont
  {Kargar}}, \bibinfo {author} {\bibfnamefont {E.~A.}\ \bibnamefont {Coleman}},
  \bibinfo {author} {\bibfnamefont {S.}~\bibnamefont {Ghosh}}, \bibinfo
  {author} {\bibfnamefont {J.}~\bibnamefont {Lee}}, \bibinfo {author}
  {\bibfnamefont {M.~J.}\ \bibnamefont {Gomez}}, \bibinfo {author}
  {\bibfnamefont {Y.}~\bibnamefont {Liu}}, \bibinfo {author} {\bibfnamefont
  {A.~S.}\ \bibnamefont {Magana}}, \bibinfo {author} {\bibfnamefont
  {Z.}~\bibnamefont {Barani}}, \bibinfo {author} {\bibfnamefont
  {A.}~\bibnamefont {Mohammadzadeh}}, \bibinfo {author} {\bibfnamefont
  {B.}~\bibnamefont {Debnath}},  \emph {et~al.},\ }\href@noop {} {\bibfield
  {journal} {\bibinfo  {journal} {ACS nano}\ }\textbf {\bibinfo {volume}
  {14}},\ \bibinfo {pages} {2424} (\bibinfo {year} {2020})}\BibitemShut
  {NoStop}%
\bibitem [{\citenamefont {Zhang}\ \emph
  {et~al.}(2021{\natexlab{a}})\citenamefont {Zhang}, \citenamefont {Hwangbo},
  \citenamefont {Wang}, \citenamefont {Jiang}, \citenamefont {Chu},
  \citenamefont {Wen}, \citenamefont {Xiao},\ and\ \citenamefont
  {Xu}}]{zhang2021observation}%
  \BibitemOpen
  \bibfield  {author} {\bibinfo {author} {\bibfnamefont {Q.}~\bibnamefont
  {Zhang}}, \bibinfo {author} {\bibfnamefont {K.}~\bibnamefont {Hwangbo}},
  \bibinfo {author} {\bibfnamefont {C.}~\bibnamefont {Wang}}, \bibinfo {author}
  {\bibfnamefont {Q.}~\bibnamefont {Jiang}}, \bibinfo {author} {\bibfnamefont
  {J.-H.}\ \bibnamefont {Chu}}, \bibinfo {author} {\bibfnamefont
  {H.}~\bibnamefont {Wen}}, \bibinfo {author} {\bibfnamefont {D.}~\bibnamefont
  {Xiao}}, \ and\ \bibinfo {author} {\bibfnamefont {X.}~\bibnamefont {Xu}},\
  }\href@noop {} {\bibfield  {journal} {\bibinfo  {journal} {Nano Letters}\
  }\textbf {\bibinfo {volume} {21}},\ \bibinfo {pages} {6938} (\bibinfo {year}
  {2021}{\natexlab{a}})}\BibitemShut {NoStop}%
\bibitem [{\citenamefont {Hwangbo}\ \emph {et~al.}(2021)\citenamefont
  {Hwangbo}, \citenamefont {Zhang}, \citenamefont {Jiang}, \citenamefont
  {Wang}, \citenamefont {Fonseca}, \citenamefont {Wang}, \citenamefont
  {Diederich}, \citenamefont {Gamelin}, \citenamefont {Xiao}, \citenamefont
  {Chu} \emph {et~al.}}]{hwangbo2021highly}%
  \BibitemOpen
  \bibfield  {author} {\bibinfo {author} {\bibfnamefont {K.}~\bibnamefont
  {Hwangbo}}, \bibinfo {author} {\bibfnamefont {Q.}~\bibnamefont {Zhang}},
  \bibinfo {author} {\bibfnamefont {Q.}~\bibnamefont {Jiang}}, \bibinfo
  {author} {\bibfnamefont {Y.}~\bibnamefont {Wang}}, \bibinfo {author}
  {\bibfnamefont {J.}~\bibnamefont {Fonseca}}, \bibinfo {author} {\bibfnamefont
  {C.}~\bibnamefont {Wang}}, \bibinfo {author} {\bibfnamefont {G.~M.}\
  \bibnamefont {Diederich}}, \bibinfo {author} {\bibfnamefont {D.~R.}\
  \bibnamefont {Gamelin}}, \bibinfo {author} {\bibfnamefont {D.}~\bibnamefont
  {Xiao}}, \bibinfo {author} {\bibfnamefont {J.-H.}\ \bibnamefont {Chu}},
  \emph {et~al.},\ }\href@noop {} {\bibfield  {journal} {\bibinfo  {journal}
  {Nature Nanotechnology}\ }\textbf {\bibinfo {volume} {16}},\ \bibinfo {pages}
  {655} (\bibinfo {year} {2021})}\BibitemShut {NoStop}%
\bibitem [{\citenamefont {Kim}\ \emph {et~al.}(2023)\citenamefont {Kim},
  \citenamefont {Huang}, \citenamefont {Guo}, \citenamefont {Li}, \citenamefont
  {Rocca}, \citenamefont {Gao}, \citenamefont {Choe}, \citenamefont {Lujan},
  \citenamefont {Wu}, \citenamefont {Lin} \emph {et~al.}}]{kim2023anisotropic}%
  \BibitemOpen
  \bibfield  {author} {\bibinfo {author} {\bibfnamefont {D.~S.}\ \bibnamefont
  {Kim}}, \bibinfo {author} {\bibfnamefont {D.}~\bibnamefont {Huang}}, \bibinfo
  {author} {\bibfnamefont {C.}~\bibnamefont {Guo}}, \bibinfo {author}
  {\bibfnamefont {K.}~\bibnamefont {Li}}, \bibinfo {author} {\bibfnamefont
  {D.}~\bibnamefont {Rocca}}, \bibinfo {author} {\bibfnamefont {F.~Y.}\
  \bibnamefont {Gao}}, \bibinfo {author} {\bibfnamefont {J.}~\bibnamefont
  {Choe}}, \bibinfo {author} {\bibfnamefont {D.}~\bibnamefont {Lujan}},
  \bibinfo {author} {\bibfnamefont {T.-H.}\ \bibnamefont {Wu}}, \bibinfo
  {author} {\bibfnamefont {K.-H.}\ \bibnamefont {Lin}},  \emph {et~al.},\
  }\href@noop {} {\bibfield  {journal} {\bibinfo  {journal} {Advanced
  Materials}\ }\textbf {\bibinfo {volume} {35}},\ \bibinfo {pages} {2206585}
  (\bibinfo {year} {2023})}\BibitemShut {NoStop}%
\bibitem [{\citenamefont {Klein}\ \emph {et~al.}(2023)\citenamefont {Klein},
  \citenamefont {Pingault}, \citenamefont {Florian}, \citenamefont
  {Hei{\ss}enbüttel}, \citenamefont {Steinhoff}, \citenamefont {Song},
  \citenamefont {Torres}, \citenamefont {Dirnberger}, \citenamefont {Curtis},
  \citenamefont {Weile} \emph {et~al.}}]{klein2023bulk}%
  \BibitemOpen
  \bibfield  {author} {\bibinfo {author} {\bibfnamefont {J.}~\bibnamefont
  {Klein}}, \bibinfo {author} {\bibfnamefont {B.}~\bibnamefont {Pingault}},
  \bibinfo {author} {\bibfnamefont {M.}~\bibnamefont {Florian}}, \bibinfo
  {author} {\bibfnamefont {M.-C.}\ \bibnamefont {Hei{\ss}enbüttel}}, \bibinfo
  {author} {\bibfnamefont {A.}~\bibnamefont {Steinhoff}}, \bibinfo {author}
  {\bibfnamefont {Z.}~\bibnamefont {Song}}, \bibinfo {author} {\bibfnamefont
  {K.}~\bibnamefont {Torres}}, \bibinfo {author} {\bibfnamefont
  {F.}~\bibnamefont {Dirnberger}}, \bibinfo {author} {\bibfnamefont {J.~B.}\
  \bibnamefont {Curtis}}, \bibinfo {author} {\bibfnamefont {M.}~\bibnamefont
  {Weile}},  \emph {et~al.},\ }\href@noop {} {\bibfield  {journal} {\bibinfo
  {journal} {ACS nano}\ }\textbf {\bibinfo {volume} {17}},\ \bibinfo {pages}
  {5316} (\bibinfo {year} {2023})}\BibitemShut {NoStop}%
\bibitem [{\citenamefont {Sun}\ \emph {et~al.}(2020)\citenamefont {Sun},
  \citenamefont {Li}, \citenamefont {Wang}, \citenamefont {Sui}, \citenamefont
  {Zhang}, \citenamefont {Wang}, \citenamefont {Liu}, \citenamefont {Li},
  \citenamefont {Feng}, \citenamefont {Zhong} \emph {et~al.}}]{sun2020room}%
  \BibitemOpen
  \bibfield  {author} {\bibinfo {author} {\bibfnamefont {X.}~\bibnamefont
  {Sun}}, \bibinfo {author} {\bibfnamefont {W.}~\bibnamefont {Li}}, \bibinfo
  {author} {\bibfnamefont {X.}~\bibnamefont {Wang}}, \bibinfo {author}
  {\bibfnamefont {Q.}~\bibnamefont {Sui}}, \bibinfo {author} {\bibfnamefont
  {T.}~\bibnamefont {Zhang}}, \bibinfo {author} {\bibfnamefont
  {Z.}~\bibnamefont {Wang}}, \bibinfo {author} {\bibfnamefont {L.}~\bibnamefont
  {Liu}}, \bibinfo {author} {\bibfnamefont {D.}~\bibnamefont {Li}}, \bibinfo
  {author} {\bibfnamefont {S.}~\bibnamefont {Feng}}, \bibinfo {author}
  {\bibfnamefont {S.}~\bibnamefont {Zhong}},  \emph {et~al.},\ }\href@noop {}
  {\bibfield  {journal} {\bibinfo  {journal} {Nano Research}\ }\textbf
  {\bibinfo {volume} {13}},\ \bibinfo {pages} {3358} (\bibinfo {year}
  {2020})}\BibitemShut {NoStop}%
\bibitem [{\citenamefont {Zhang}\ \emph
  {et~al.}(2021{\natexlab{b}})\citenamefont {Zhang}, \citenamefont {Lu},
  \citenamefont {Liu}, \citenamefont {Niu}, \citenamefont {Sun}, \citenamefont
  {Cook}, \citenamefont {Vaninger}, \citenamefont {Miceli}, \citenamefont
  {Singh}, \citenamefont {Lian} \emph {et~al.}}]{zhang2021room}%
  \BibitemOpen
  \bibfield  {author} {\bibinfo {author} {\bibfnamefont {X.}~\bibnamefont
  {Zhang}}, \bibinfo {author} {\bibfnamefont {Q.}~\bibnamefont {Lu}}, \bibinfo
  {author} {\bibfnamefont {W.}~\bibnamefont {Liu}}, \bibinfo {author}
  {\bibfnamefont {W.}~\bibnamefont {Niu}}, \bibinfo {author} {\bibfnamefont
  {J.}~\bibnamefont {Sun}}, \bibinfo {author} {\bibfnamefont {J.}~\bibnamefont
  {Cook}}, \bibinfo {author} {\bibfnamefont {M.}~\bibnamefont {Vaninger}},
  \bibinfo {author} {\bibfnamefont {P.~F.}\ \bibnamefont {Miceli}}, \bibinfo
  {author} {\bibfnamefont {D.~J.}\ \bibnamefont {Singh}}, \bibinfo {author}
  {\bibfnamefont {S.-W.}\ \bibnamefont {Lian}},  \emph {et~al.},\ }\href@noop
  {} {\bibfield  {journal} {\bibinfo  {journal} {Nature communications}\
  }\textbf {\bibinfo {volume} {12}},\ \bibinfo {pages} {1} (\bibinfo {year}
  {2021}{\natexlab{b}})}\BibitemShut {NoStop}%
\bibitem [{\citenamefont {Xian}\ \emph {et~al.}(2022)\citenamefont {Xian},
  \citenamefont {Wang}, \citenamefont {Nie}, \citenamefont {Li}, \citenamefont
  {Han}, \citenamefont {Lin}, \citenamefont {Zhang}, \citenamefont {Liu},
  \citenamefont {Zhang}, \citenamefont {Miao} \emph {et~al.}}]{xian2022spin}%
  \BibitemOpen
  \bibfield  {author} {\bibinfo {author} {\bibfnamefont {J.-J.}\ \bibnamefont
  {Xian}}, \bibinfo {author} {\bibfnamefont {C.}~\bibnamefont {Wang}}, \bibinfo
  {author} {\bibfnamefont {J.-H.}\ \bibnamefont {Nie}}, \bibinfo {author}
  {\bibfnamefont {R.}~\bibnamefont {Li}}, \bibinfo {author} {\bibfnamefont
  {M.}~\bibnamefont {Han}}, \bibinfo {author} {\bibfnamefont {J.}~\bibnamefont
  {Lin}}, \bibinfo {author} {\bibfnamefont {W.-H.}\ \bibnamefont {Zhang}},
  \bibinfo {author} {\bibfnamefont {Z.-Y.}\ \bibnamefont {Liu}}, \bibinfo
  {author} {\bibfnamefont {Z.-M.}\ \bibnamefont {Zhang}}, \bibinfo {author}
  {\bibfnamefont {M.-P.}\ \bibnamefont {Miao}},  \emph {et~al.},\ }\href@noop
  {} {\bibfield  {journal} {\bibinfo  {journal} {Nature Communications}\
  }\textbf {\bibinfo {volume} {13}},\ \bibinfo {pages} {1} (\bibinfo {year}
  {2022})}\BibitemShut {NoStop}%
\bibitem [{\citenamefont {Lv}\ \emph {et~al.}(2015)\citenamefont {Lv},
  \citenamefont {Lu}, \citenamefont {Shao}, \citenamefont {Liu},\ and\
  \citenamefont {Sun}}]{lv2015strain}%
  \BibitemOpen
  \bibfield  {author} {\bibinfo {author} {\bibfnamefont {H.}~\bibnamefont
  {Lv}}, \bibinfo {author} {\bibfnamefont {W.}~\bibnamefont {Lu}}, \bibinfo
  {author} {\bibfnamefont {D.}~\bibnamefont {Shao}}, \bibinfo {author}
  {\bibfnamefont {Y.}~\bibnamefont {Liu}}, \ and\ \bibinfo {author}
  {\bibfnamefont {Y.}~\bibnamefont {Sun}},\ }\href@noop {} {\bibfield
  {journal} {\bibinfo  {journal} {Physical Review B}\ }\textbf {\bibinfo
  {volume} {92}},\ \bibinfo {pages} {214419} (\bibinfo {year}
  {2015})}\BibitemShut {NoStop}%
\bibitem [{\citenamefont {Zhou}\ \emph {et~al.}(2022)\citenamefont {Zhou},
  \citenamefont {Song}, \citenamefont {Chai}, \citenamefont {Wong},
  \citenamefont {Xu}, \citenamefont {Jiang}, \citenamefont {Feng},
  \citenamefont {Yang},\ and\ \citenamefont {Wang}}]{zhou2022structure}%
  \BibitemOpen
  \bibfield  {author} {\bibinfo {author} {\bibfnamefont {J.}~\bibnamefont
  {Zhou}}, \bibinfo {author} {\bibfnamefont {X.}~\bibnamefont {Song}}, \bibinfo
  {author} {\bibfnamefont {J.}~\bibnamefont {Chai}}, \bibinfo {author}
  {\bibfnamefont {N.~L.~M.}\ \bibnamefont {Wong}}, \bibinfo {author}
  {\bibfnamefont {X.}~\bibnamefont {Xu}}, \bibinfo {author} {\bibfnamefont
  {Y.}~\bibnamefont {Jiang}}, \bibinfo {author} {\bibfnamefont {Y.~P.}\
  \bibnamefont {Feng}}, \bibinfo {author} {\bibfnamefont {M.}~\bibnamefont
  {Yang}}, \ and\ \bibinfo {author} {\bibfnamefont {S.}~\bibnamefont {Wang}},\
  }\href@noop {} {\bibfield  {journal} {\bibinfo  {journal} {Journal of Alloys
  and Compounds}\ }\textbf {\bibinfo {volume} {893}},\ \bibinfo {pages}
  {162223} (\bibinfo {year} {2022})}\BibitemShut {NoStop}%
\bibitem [{\citenamefont {Gao}\ \emph {et~al.}(2021)\citenamefont {Gao},
  \citenamefont {Li},\ and\ \citenamefont {Yang}}]{gao2021thickness}%
  \BibitemOpen
  \bibfield  {author} {\bibinfo {author} {\bibfnamefont {P.}~\bibnamefont
  {Gao}}, \bibinfo {author} {\bibfnamefont {X.}~\bibnamefont {Li}}, \ and\
  \bibinfo {author} {\bibfnamefont {J.}~\bibnamefont {Yang}},\ }\href@noop {}
  {\bibfield  {journal} {\bibinfo  {journal} {The Journal of Physical Chemistry
  Letters}\ }\textbf {\bibinfo {volume} {12}},\ \bibinfo {pages} {6847}
  (\bibinfo {year} {2021})}\BibitemShut {NoStop}%
\bibitem [{\citenamefont {Otero~Fumega}\ \emph {et~al.}(2020)\citenamefont
  {Otero~Fumega}, \citenamefont {Phillips},\ and\ \citenamefont
  {Pardo}}]{otero2020controlled}%
  \BibitemOpen
  \bibfield  {author} {\bibinfo {author} {\bibfnamefont {A.}~\bibnamefont
  {Otero~Fumega}}, \bibinfo {author} {\bibfnamefont {J.}~\bibnamefont
  {Phillips}}, \ and\ \bibinfo {author} {\bibfnamefont {V.}~\bibnamefont
  {Pardo}},\ }\href@noop {} {\bibfield  {journal} {\bibinfo  {journal} {The
  Journal of Physical Chemistry C}\ }\textbf {\bibinfo {volume} {124}},\
  \bibinfo {pages} {21047} (\bibinfo {year} {2020})}\BibitemShut {NoStop}%
\bibitem [{\citenamefont {Wilson}\ \emph {et~al.}(1975)\citenamefont {Wilson},
  \citenamefont {Di~Salvo},\ and\ \citenamefont {Mahajan}}]{wilson1975charge}%
  \BibitemOpen
  \bibfield  {author} {\bibinfo {author} {\bibfnamefont {J.~A.}\ \bibnamefont
  {Wilson}}, \bibinfo {author} {\bibfnamefont {F.}~\bibnamefont {Di~Salvo}}, \
  and\ \bibinfo {author} {\bibfnamefont {S.}~\bibnamefont {Mahajan}},\
  }\href@noop {} {\bibfield  {journal} {\bibinfo  {journal} {Advances in
  Physics}\ }\textbf {\bibinfo {volume} {24}},\ \bibinfo {pages} {117}
  (\bibinfo {year} {1975})}\BibitemShut {NoStop}%
\bibitem [{\citenamefont {Wilson}\ \emph {et~al.}(1974)\citenamefont {Wilson},
  \citenamefont {Di~Salvo},\ and\ \citenamefont {Mahajan}}]{wilson1974charge}%
  \BibitemOpen
  \bibfield  {author} {\bibinfo {author} {\bibfnamefont {J.}~\bibnamefont
  {Wilson}}, \bibinfo {author} {\bibfnamefont {F.}~\bibnamefont {Di~Salvo}}, \
  and\ \bibinfo {author} {\bibfnamefont {S.}~\bibnamefont {Mahajan}},\
  }\href@noop {} {\bibfield  {journal} {\bibinfo  {journal} {Physical review
  letters}\ }\textbf {\bibinfo {volume} {32}},\ \bibinfo {pages} {882}
  (\bibinfo {year} {1974})}\BibitemShut {NoStop}%
\bibitem [{\citenamefont {Rossnagel}(2011)}]{rossnagel2011origin}%
  \BibitemOpen
  \bibfield  {author} {\bibinfo {author} {\bibfnamefont {K.}~\bibnamefont
  {Rossnagel}},\ }\href@noop {} {\bibfield  {journal} {\bibinfo  {journal}
  {Journal of Physics: Condensed Matter}\ }\textbf {\bibinfo {volume} {23}},\
  \bibinfo {pages} {213001} (\bibinfo {year} {2011})}\BibitemShut {NoStop}%
\bibitem [{\citenamefont {GRUNER}(1994)}]{gruner1994density}%
  \BibitemOpen
  \bibfield  {author} {\bibinfo {author} {\bibfnamefont {G.}~\bibnamefont
  {GRUNER}},\ }\href@noop {} {\bibfield  {journal} {\bibinfo  {journal}
  {Frontiers In Physics}\ }\textbf {\bibinfo {volume} {89}} (\bibinfo {year}
  {1994})}\BibitemShut {NoStop}%
\bibitem [{\citenamefont {Hu}\ \emph {et~al.}(2022)\citenamefont {Hu},
  \citenamefont {Zhang}, \citenamefont {Zhao}, \citenamefont {Chen},
  \citenamefont {Ding}, \citenamefont {Yang}, \citenamefont {Wang},
  \citenamefont {Li}, \citenamefont {Wang}, \citenamefont {Feng} \emph
  {et~al.}}]{hu2022real}%
  \BibitemOpen
  \bibfield  {author} {\bibinfo {author} {\bibfnamefont {Y.}~\bibnamefont
  {Hu}}, \bibinfo {author} {\bibfnamefont {T.}~\bibnamefont {Zhang}}, \bibinfo
  {author} {\bibfnamefont {D.}~\bibnamefont {Zhao}}, \bibinfo {author}
  {\bibfnamefont {C.}~\bibnamefont {Chen}}, \bibinfo {author} {\bibfnamefont
  {S.}~\bibnamefont {Ding}}, \bibinfo {author} {\bibfnamefont {W.}~\bibnamefont
  {Yang}}, \bibinfo {author} {\bibfnamefont {X.}~\bibnamefont {Wang}}, \bibinfo
  {author} {\bibfnamefont {C.}~\bibnamefont {Li}}, \bibinfo {author}
  {\bibfnamefont {H.}~\bibnamefont {Wang}}, \bibinfo {author} {\bibfnamefont
  {D.}~\bibnamefont {Feng}},  \emph {et~al.},\ }\href@noop {} {\bibfield
  {journal} {\bibinfo  {journal} {Nature Communications}\ }\textbf {\bibinfo
  {volume} {13}},\ \bibinfo {pages} {1} (\bibinfo {year} {2022})}\BibitemShut
  {NoStop}%
\bibitem [{\citenamefont {Kresse}\ and\ \citenamefont
  {Furthm{\"u}ller}(1996)}]{kresse1996efficient}%
  \BibitemOpen
  \bibfield  {author} {\bibinfo {author} {\bibfnamefont {G.}~\bibnamefont
  {Kresse}}\ and\ \bibinfo {author} {\bibfnamefont {J.}~\bibnamefont
  {Furthm{\"u}ller}},\ }\href@noop {} {\bibfield  {journal} {\bibinfo
  {journal} {Physical review B}\ }\textbf {\bibinfo {volume} {54}},\ \bibinfo
  {pages} {11169} (\bibinfo {year} {1996})}\BibitemShut {NoStop}%
\bibitem [{\citenamefont {Kresse}\ and\ \citenamefont
  {Joubert}(1999)}]{kresse1999ultrasoft}%
  \BibitemOpen
  \bibfield  {author} {\bibinfo {author} {\bibfnamefont {G.}~\bibnamefont
  {Kresse}}\ and\ \bibinfo {author} {\bibfnamefont {D.}~\bibnamefont
  {Joubert}},\ }\href@noop {} {\bibfield  {journal} {\bibinfo  {journal}
  {Physical review b}\ }\textbf {\bibinfo {volume} {59}},\ \bibinfo {pages}
  {1758} (\bibinfo {year} {1999})}\BibitemShut {NoStop}%
\bibitem [{\citenamefont {Perdew}\ \emph {et~al.}(1996)\citenamefont {Perdew},
  \citenamefont {Burke},\ and\ \citenamefont
  {Ernzerhof}}]{perdew1996generalized}%
  \BibitemOpen
  \bibfield  {author} {\bibinfo {author} {\bibfnamefont {J.~P.}\ \bibnamefont
  {Perdew}}, \bibinfo {author} {\bibfnamefont {K.}~\bibnamefont {Burke}}, \
  and\ \bibinfo {author} {\bibfnamefont {M.}~\bibnamefont {Ernzerhof}},\
  }\href@noop {} {\bibfield  {journal} {\bibinfo  {journal} {Physical review
  letters}\ }\textbf {\bibinfo {volume} {77}},\ \bibinfo {pages} {3865}
  (\bibinfo {year} {1996})}\BibitemShut {NoStop}%
\bibitem [{\citenamefont {Monkhorst}\ and\ \citenamefont
  {Pack}(1976)}]{monkhorst1976special}%
  \BibitemOpen
  \bibfield  {author} {\bibinfo {author} {\bibfnamefont {H.~J.}\ \bibnamefont
  {Monkhorst}}\ and\ \bibinfo {author} {\bibfnamefont {J.~D.}\ \bibnamefont
  {Pack}},\ }\href@noop {} {\bibfield  {journal} {\bibinfo  {journal} {Physical
  review B}\ }\textbf {\bibinfo {volume} {13}},\ \bibinfo {pages} {5188}
  (\bibinfo {year} {1976})}\BibitemShut {NoStop}%
\bibitem [{\citenamefont {Liu}\ \emph {et~al.}(2022)\citenamefont {Liu},
  \citenamefont {Kwon}, \citenamefont {de~Coster}, \citenamefont {Lake},\ and\
  \citenamefont {Neupane}}]{liu2022structural}%
  \BibitemOpen
  \bibfield  {author} {\bibinfo {author} {\bibfnamefont {Y.}~\bibnamefont
  {Liu}}, \bibinfo {author} {\bibfnamefont {S.}~\bibnamefont {Kwon}}, \bibinfo
  {author} {\bibfnamefont {G.~J.}\ \bibnamefont {de~Coster}}, \bibinfo {author}
  {\bibfnamefont {R.~K.}\ \bibnamefont {Lake}}, \ and\ \bibinfo {author}
  {\bibfnamefont {M.~R.}\ \bibnamefont {Neupane}},\ }\href@noop {} {\bibfield
  {journal} {\bibinfo  {journal} {Physical Review Materials}\ }\textbf
  {\bibinfo {volume} {6}},\ \bibinfo {pages} {084004} (\bibinfo {year}
  {2022})}\BibitemShut {NoStop}%
\bibitem [{\citenamefont {Rauch}\ \emph
  {et~al.}(2020{\natexlab{a}})\citenamefont {Rauch}, \citenamefont {Marques},\
  and\ \citenamefont {Botti}}]{rauch2020local}%
  \BibitemOpen
  \bibfield  {author} {\bibinfo {author} {\bibfnamefont {T.}~\bibnamefont
  {Rauch}}, \bibinfo {author} {\bibfnamefont {M.~A.}\ \bibnamefont {Marques}},
  \ and\ \bibinfo {author} {\bibfnamefont {S.}~\bibnamefont {Botti}},\
  }\href@noop {} {\bibfield  {journal} {\bibinfo  {journal} {Journal of
  chemical theory and computation}\ }\textbf {\bibinfo {volume} {16}},\
  \bibinfo {pages} {2654} (\bibinfo {year} {2020}{\natexlab{a}})}\BibitemShut
  {NoStop}%
\bibitem [{\citenamefont {Tran}\ \emph {et~al.}(2021)\citenamefont {Tran},
  \citenamefont {Doumont}, \citenamefont {Kalantari}, \citenamefont {Blaha},
  \citenamefont {Rauch}, \citenamefont {Borlido}, \citenamefont {Botti},
  \citenamefont {Marques}, \citenamefont {Patra}, \citenamefont {Jana} \emph
  {et~al.}}]{tran2021bandgap}%
  \BibitemOpen
  \bibfield  {author} {\bibinfo {author} {\bibfnamefont {F.}~\bibnamefont
  {Tran}}, \bibinfo {author} {\bibfnamefont {J.}~\bibnamefont {Doumont}},
  \bibinfo {author} {\bibfnamefont {L.}~\bibnamefont {Kalantari}}, \bibinfo
  {author} {\bibfnamefont {P.}~\bibnamefont {Blaha}}, \bibinfo {author}
  {\bibfnamefont {T.}~\bibnamefont {Rauch}}, \bibinfo {author} {\bibfnamefont
  {P.}~\bibnamefont {Borlido}}, \bibinfo {author} {\bibfnamefont
  {S.}~\bibnamefont {Botti}}, \bibinfo {author} {\bibfnamefont {M.~A.}\
  \bibnamefont {Marques}}, \bibinfo {author} {\bibfnamefont {A.}~\bibnamefont
  {Patra}}, \bibinfo {author} {\bibfnamefont {S.}~\bibnamefont {Jana}},  \emph
  {et~al.},\ }\href@noop {} {\bibfield  {journal} {\bibinfo  {journal} {The
  Journal of Chemical Physics}\ }\textbf {\bibinfo {volume} {155}},\ \bibinfo
  {pages} {104103} (\bibinfo {year} {2021})}\BibitemShut {NoStop}%
\bibitem [{\citenamefont {Rauch}\ \emph
  {et~al.}(2020{\natexlab{b}})\citenamefont {Rauch}, \citenamefont {Marques},\
  and\ \citenamefont {Botti}}]{rauch2020accurate}%
  \BibitemOpen
  \bibfield  {author} {\bibinfo {author} {\bibfnamefont {T.}~\bibnamefont
  {Rauch}}, \bibinfo {author} {\bibfnamefont {M.~A.}\ \bibnamefont {Marques}},
  \ and\ \bibinfo {author} {\bibfnamefont {S.}~\bibnamefont {Botti}},\
  }\href@noop {} {\bibfield  {journal} {\bibinfo  {journal} {Physical Review
  B}\ }\textbf {\bibinfo {volume} {101}},\ \bibinfo {pages} {245163} (\bibinfo
  {year} {2020}{\natexlab{b}})}\BibitemShut {NoStop}%
\bibitem [{\citenamefont {Freitas}\ \emph {et~al.}(2015)\citenamefont
  {Freitas}, \citenamefont {Weht}, \citenamefont {Sulpice}, \citenamefont
  {Remenyi}, \citenamefont {Strobel}, \citenamefont {Gay}, \citenamefont
  {Marcus},\ and\ \citenamefont
  {N{\'u}{\~n}ez-Regueiro}}]{freitas2015ferromagnetism}%
  \BibitemOpen
  \bibfield  {author} {\bibinfo {author} {\bibfnamefont {D.~C.}\ \bibnamefont
  {Freitas}}, \bibinfo {author} {\bibfnamefont {R.}~\bibnamefont {Weht}},
  \bibinfo {author} {\bibfnamefont {A.}~\bibnamefont {Sulpice}}, \bibinfo
  {author} {\bibfnamefont {G.}~\bibnamefont {Remenyi}}, \bibinfo {author}
  {\bibfnamefont {P.}~\bibnamefont {Strobel}}, \bibinfo {author} {\bibfnamefont
  {F.}~\bibnamefont {Gay}}, \bibinfo {author} {\bibfnamefont {J.}~\bibnamefont
  {Marcus}}, \ and\ \bibinfo {author} {\bibfnamefont {M.}~\bibnamefont
  {N{\'u}{\~n}ez-Regueiro}},\ }\href@noop {} {\bibfield  {journal} {\bibinfo
  {journal} {Journal of Physics: Condensed Matter}\ }\textbf {\bibinfo {volume}
  {27}},\ \bibinfo {pages} {176002} (\bibinfo {year} {2015})}\BibitemShut
  {NoStop}%
\bibitem [{\citenamefont {Sun}\ \emph {et~al.}(2021)\citenamefont {Sun},
  \citenamefont {Yan}, \citenamefont {Ning}, \citenamefont {Zhang},
  \citenamefont {Zhao}, \citenamefont {Gao}, \citenamefont {Kanagaraj},
  \citenamefont {Zhang}, \citenamefont {Li}, \citenamefont {Lu} \emph
  {et~al.}}]{sun2021ferromagnetism}%
  \BibitemOpen
  \bibfield  {author} {\bibinfo {author} {\bibfnamefont {Y.}~\bibnamefont
  {Sun}}, \bibinfo {author} {\bibfnamefont {P.}~\bibnamefont {Yan}}, \bibinfo
  {author} {\bibfnamefont {J.}~\bibnamefont {Ning}}, \bibinfo {author}
  {\bibfnamefont {X.}~\bibnamefont {Zhang}}, \bibinfo {author} {\bibfnamefont
  {Y.}~\bibnamefont {Zhao}}, \bibinfo {author} {\bibfnamefont {Q.}~\bibnamefont
  {Gao}}, \bibinfo {author} {\bibfnamefont {M.}~\bibnamefont {Kanagaraj}},
  \bibinfo {author} {\bibfnamefont {K.}~\bibnamefont {Zhang}}, \bibinfo
  {author} {\bibfnamefont {J.}~\bibnamefont {Li}}, \bibinfo {author}
  {\bibfnamefont {X.}~\bibnamefont {Lu}},  \emph {et~al.},\ }\href@noop {}
  {\bibfield  {journal} {\bibinfo  {journal} {AIP Advances}\ }\textbf {\bibinfo
  {volume} {11}},\ \bibinfo {pages} {035138} (\bibinfo {year}
  {2021})}\BibitemShut {NoStop}%
\bibitem [{\citenamefont {Meng}\ \emph {et~al.}(2021)\citenamefont {Meng},
  \citenamefont {Zhou}, \citenamefont {Xu}, \citenamefont {Yang}, \citenamefont
  {Si}, \citenamefont {Liu}, \citenamefont {Wang}, \citenamefont {Jiang},
  \citenamefont {Li}, \citenamefont {Qin} \emph {et~al.}}]{meng2021anomalous}%
  \BibitemOpen
  \bibfield  {author} {\bibinfo {author} {\bibfnamefont {L.}~\bibnamefont
  {Meng}}, \bibinfo {author} {\bibfnamefont {Z.}~\bibnamefont {Zhou}}, \bibinfo
  {author} {\bibfnamefont {M.}~\bibnamefont {Xu}}, \bibinfo {author}
  {\bibfnamefont {S.}~\bibnamefont {Yang}}, \bibinfo {author} {\bibfnamefont
  {K.}~\bibnamefont {Si}}, \bibinfo {author} {\bibfnamefont {L.}~\bibnamefont
  {Liu}}, \bibinfo {author} {\bibfnamefont {X.}~\bibnamefont {Wang}}, \bibinfo
  {author} {\bibfnamefont {H.}~\bibnamefont {Jiang}}, \bibinfo {author}
  {\bibfnamefont {B.}~\bibnamefont {Li}}, \bibinfo {author} {\bibfnamefont
  {P.}~\bibnamefont {Qin}},  \emph {et~al.},\ }\href@noop {} {\bibfield
  {journal} {\bibinfo  {journal} {Nature communications}\ }\textbf {\bibinfo
  {volume} {12}},\ \bibinfo {pages} {1} (\bibinfo {year} {2021})}\BibitemShut
  {NoStop}%
\bibitem [{\citenamefont {Webster}\ and\ \citenamefont
  {Yan}(2018)}]{webster2018strain}%
  \BibitemOpen
  \bibfield  {author} {\bibinfo {author} {\bibfnamefont {L.}~\bibnamefont
  {Webster}}\ and\ \bibinfo {author} {\bibfnamefont {J.-A.}\ \bibnamefont
  {Yan}},\ }\href@noop {} {\bibfield  {journal} {\bibinfo  {journal} {Physical
  Review B}\ }\textbf {\bibinfo {volume} {98}},\ \bibinfo {pages} {144411}
  (\bibinfo {year} {2018})}\BibitemShut {NoStop}%
\bibitem [{\citenamefont {Hu}\ \emph {et~al.}(2020)\citenamefont {Hu},
  \citenamefont {Zhao}, \citenamefont {Shen}, \citenamefont {Krasheninnikov},
  \citenamefont {Chen},\ and\ \citenamefont {Sun}}]{hu2020enhanced}%
  \BibitemOpen
  \bibfield  {author} {\bibinfo {author} {\bibfnamefont {X.}~\bibnamefont
  {Hu}}, \bibinfo {author} {\bibfnamefont {Y.}~\bibnamefont {Zhao}}, \bibinfo
  {author} {\bibfnamefont {X.}~\bibnamefont {Shen}}, \bibinfo {author}
  {\bibfnamefont {A.~V.}\ \bibnamefont {Krasheninnikov}}, \bibinfo {author}
  {\bibfnamefont {Z.}~\bibnamefont {Chen}}, \ and\ \bibinfo {author}
  {\bibfnamefont {L.}~\bibnamefont {Sun}},\ }\href@noop {} {\bibfield
  {journal} {\bibinfo  {journal} {ACS applied materials \& interfaces}\
  }\textbf {\bibinfo {volume} {12}},\ \bibinfo {pages} {26367} (\bibinfo {year}
  {2020})}\BibitemShut {NoStop}%
\bibitem [{\citenamefont {Bergeron}\ \emph {et~al.}(2012)\citenamefont
  {Bergeron}, \citenamefont {Chowdhury}, \citenamefont {Punk}, \citenamefont
  {Sachdev},\ and\ \citenamefont {Tremblay}}]{bergeron2012breakdown}%
  \BibitemOpen
  \bibfield  {author} {\bibinfo {author} {\bibfnamefont {D.}~\bibnamefont
  {Bergeron}}, \bibinfo {author} {\bibfnamefont {D.}~\bibnamefont {Chowdhury}},
  \bibinfo {author} {\bibfnamefont {M.}~\bibnamefont {Punk}}, \bibinfo {author}
  {\bibfnamefont {S.}~\bibnamefont {Sachdev}}, \ and\ \bibinfo {author}
  {\bibfnamefont {A.-M.}\ \bibnamefont {Tremblay}},\ }\href@noop {} {\bibfield
  {journal} {\bibinfo  {journal} {Physical Review B}\ }\textbf {\bibinfo
  {volume} {86}},\ \bibinfo {pages} {155123} (\bibinfo {year}
  {2012})}\BibitemShut {NoStop}%
\bibitem [{\citenamefont {Sachdev}\ \emph {et~al.}(2016)\citenamefont
  {Sachdev}, \citenamefont {Berg}, \citenamefont {Chatterjee},\ and\
  \citenamefont {Schattner}}]{sachdev2016spin}%
  \BibitemOpen
  \bibfield  {author} {\bibinfo {author} {\bibfnamefont {S.}~\bibnamefont
  {Sachdev}}, \bibinfo {author} {\bibfnamefont {E.}~\bibnamefont {Berg}},
  \bibinfo {author} {\bibfnamefont {S.}~\bibnamefont {Chatterjee}}, \ and\
  \bibinfo {author} {\bibfnamefont {Y.}~\bibnamefont {Schattner}},\ }\href@noop
  {} {\bibfield  {journal} {\bibinfo  {journal} {Physical Review B}\ }\textbf
  {\bibinfo {volume} {94}},\ \bibinfo {pages} {115147} (\bibinfo {year}
  {2016})}\BibitemShut {NoStop}%
\bibitem [{\citenamefont {S{\`y}kora}\ and\ \citenamefont
  {Metzner}(2021)}]{sykora2021fluctuation}%
  \BibitemOpen
  \bibfield  {author} {\bibinfo {author} {\bibfnamefont {J.}~\bibnamefont
  {S{\`y}kora}}\ and\ \bibinfo {author} {\bibfnamefont {W.}~\bibnamefont
  {Metzner}},\ }\href@noop {} {\bibfield  {journal} {\bibinfo  {journal}
  {Physical Review B}\ }\textbf {\bibinfo {volume} {104}},\ \bibinfo {pages}
  {125123} (\bibinfo {year} {2021})}\BibitemShut {NoStop}%
\bibitem [{\citenamefont {Sachdev}\ \emph {et~al.}(2012)\citenamefont
  {Sachdev}, \citenamefont {Metlitski},\ and\ \citenamefont
  {Punk}}]{sachdev2012antiferromagnetism}%
  \BibitemOpen
  \bibfield  {author} {\bibinfo {author} {\bibfnamefont {S.}~\bibnamefont
  {Sachdev}}, \bibinfo {author} {\bibfnamefont {M.~A.}\ \bibnamefont
  {Metlitski}}, \ and\ \bibinfo {author} {\bibfnamefont {M.}~\bibnamefont
  {Punk}},\ }\href@noop {} {\bibfield  {journal} {\bibinfo  {journal} {Journal
  of Physics: Condensed Matter}\ }\textbf {\bibinfo {volume} {24}},\ \bibinfo
  {pages} {294205} (\bibinfo {year} {2012})}\BibitemShut {NoStop}%
\bibitem [{\citenamefont {K{\"u}bler}(1980)}]{kubler1980spin}%
  \BibitemOpen
  \bibfield  {author} {\bibinfo {author} {\bibfnamefont {J.}~\bibnamefont
  {K{\"u}bler}},\ }\href@noop {} {\bibfield  {journal} {\bibinfo  {journal}
  {Journal of Magnetism and Magnetic Materials}\ }\textbf {\bibinfo {volume}
  {20}},\ \bibinfo {pages} {277} (\bibinfo {year} {1980})}\BibitemShut
  {NoStop}%
\bibitem [{\citenamefont {Fawcett}\ \emph {et~al.}(1994)\citenamefont
  {Fawcett}, \citenamefont {Alberts}, \citenamefont {Galkin}, \citenamefont
  {Noakes},\ and\ \citenamefont {Yakhmi}}]{fawcett1994spin}%
  \BibitemOpen
  \bibfield  {author} {\bibinfo {author} {\bibfnamefont {E.}~\bibnamefont
  {Fawcett}}, \bibinfo {author} {\bibfnamefont {H.}~\bibnamefont {Alberts}},
  \bibinfo {author} {\bibfnamefont {V.~Y.}\ \bibnamefont {Galkin}}, \bibinfo
  {author} {\bibfnamefont {D.}~\bibnamefont {Noakes}}, \ and\ \bibinfo {author}
  {\bibfnamefont {J.}~\bibnamefont {Yakhmi}},\ }\href@noop {} {\bibfield
  {journal} {\bibinfo  {journal} {Reviews of modern physics}\ }\textbf
  {\bibinfo {volume} {66}},\ \bibinfo {pages} {25} (\bibinfo {year}
  {1994})}\BibitemShut {NoStop}%
\bibitem [{\citenamefont {Fawcett}(1988)}]{fawcett1988spin}%
  \BibitemOpen
  \bibfield  {author} {\bibinfo {author} {\bibfnamefont {E.}~\bibnamefont
  {Fawcett}},\ }\href@noop {} {\bibfield  {journal} {\bibinfo  {journal}
  {Reviews of Modern Physics}\ }\textbf {\bibinfo {volume} {60}},\ \bibinfo
  {pages} {209} (\bibinfo {year} {1988})}\BibitemShut {NoStop}%
\bibitem [{\citenamefont {Niklasson}\ \emph {et~al.}(1999)\citenamefont
  {Niklasson}, \citenamefont {Johansson},\ and\ \citenamefont
  {Nordstr{\"o}m}}]{niklasson1999spin}%
  \BibitemOpen
  \bibfield  {author} {\bibinfo {author} {\bibfnamefont {A.}~\bibnamefont
  {Niklasson}}, \bibinfo {author} {\bibfnamefont {B.}~\bibnamefont
  {Johansson}}, \ and\ \bibinfo {author} {\bibfnamefont {L.}~\bibnamefont
  {Nordstr{\"o}m}},\ }\href@noop {} {\bibfield  {journal} {\bibinfo  {journal}
  {Physical review letters}\ }\textbf {\bibinfo {volume} {82}},\ \bibinfo
  {pages} {4544} (\bibinfo {year} {1999})}\BibitemShut {NoStop}%
\bibitem [{\citenamefont {Shibatani}\ \emph {et~al.}(1969)\citenamefont
  {Shibatani}, \citenamefont {Motizuki},\ and\ \citenamefont
  {Nagamiya}}]{shibatani1969spin}%
  \BibitemOpen
  \bibfield  {author} {\bibinfo {author} {\bibfnamefont {A.}~\bibnamefont
  {Shibatani}}, \bibinfo {author} {\bibfnamefont {K.}~\bibnamefont {Motizuki}},
  \ and\ \bibinfo {author} {\bibfnamefont {T.}~\bibnamefont {Nagamiya}},\
  }\href@noop {} {\bibfield  {journal} {\bibinfo  {journal} {Physical Review}\
  }\textbf {\bibinfo {volume} {177}},\ \bibinfo {pages} {984} (\bibinfo {year}
  {1969})}\BibitemShut {NoStop}%
\bibitem [{\citenamefont {Hirai}(1997)}]{hirai1997magnetism}%
  \BibitemOpen
  \bibfield  {author} {\bibinfo {author} {\bibfnamefont {K.}~\bibnamefont
  {Hirai}},\ }\href@noop {} {\bibfield  {journal} {\bibinfo  {journal} {Journal
  of the Physical Society of Japan}\ }\textbf {\bibinfo {volume} {66}},\
  \bibinfo {pages} {560} (\bibinfo {year} {1997})}\BibitemShut {NoStop}%
\bibitem [{\citenamefont {Kulikov}\ and\ \citenamefont
  {Tugushev}(1984)}]{kulikov1984spin}%
  \BibitemOpen
  \bibfield  {author} {\bibinfo {author} {\bibfnamefont {N.}~\bibnamefont
  {Kulikov}}\ and\ \bibinfo {author} {\bibfnamefont {V.~V.}\ \bibnamefont
  {Tugushev}},\ }\href@noop {} {\bibfield  {journal} {\bibinfo  {journal}
  {Soviet Physics Uspekhi}\ }\textbf {\bibinfo {volume} {27}},\ \bibinfo
  {pages} {954} (\bibinfo {year} {1984})}\BibitemShut {NoStop}%
\bibitem [{\citenamefont {Mannix}\ \emph {et~al.}(2001)\citenamefont {Mannix},
  \citenamefont {de~Camargo}, \citenamefont {Giles}, \citenamefont
  {de~Oliveira}, \citenamefont {Yokaichiya},\ and\ \citenamefont
  {Vettier}}]{mannix2001chromium}%
  \BibitemOpen
  \bibfield  {author} {\bibinfo {author} {\bibfnamefont {D.}~\bibnamefont
  {Mannix}}, \bibinfo {author} {\bibfnamefont {P.}~\bibnamefont {de~Camargo}},
  \bibinfo {author} {\bibfnamefont {C.}~\bibnamefont {Giles}}, \bibinfo
  {author} {\bibfnamefont {A.}~\bibnamefont {de~Oliveira}}, \bibinfo {author}
  {\bibfnamefont {F.}~\bibnamefont {Yokaichiya}}, \ and\ \bibinfo {author}
  {\bibfnamefont {C.}~\bibnamefont {Vettier}},\ }\href@noop {} {\bibfield
  {journal} {\bibinfo  {journal} {The European Physical Journal B-Condensed
  Matter and Complex Systems}\ }\textbf {\bibinfo {volume} {20}},\ \bibinfo
  {pages} {19} (\bibinfo {year} {2001})}\BibitemShut {NoStop}%
\bibitem [{\citenamefont {Chowdhury}\ \emph {et~al.}(2022)\citenamefont
  {Chowdhury}, \citenamefont {Rigosi}, \citenamefont {Hill}, \citenamefont
  {Vora}, \citenamefont {Hight~Walker},\ and\ \citenamefont
  {Tavazza}}]{chowdhury2022computational}%
  \BibitemOpen
  \bibfield  {author} {\bibinfo {author} {\bibfnamefont {S.}~\bibnamefont
  {Chowdhury}}, \bibinfo {author} {\bibfnamefont {A.~F.}\ \bibnamefont
  {Rigosi}}, \bibinfo {author} {\bibfnamefont {H.~M.}\ \bibnamefont {Hill}},
  \bibinfo {author} {\bibfnamefont {P.}~\bibnamefont {Vora}}, \bibinfo {author}
  {\bibfnamefont {A.~R.}\ \bibnamefont {Hight~Walker}}, \ and\ \bibinfo
  {author} {\bibfnamefont {F.}~\bibnamefont {Tavazza}},\ }\href@noop {}
  {\bibfield  {journal} {\bibinfo  {journal} {Nanomaterials}\ }\textbf
  {\bibinfo {volume} {12}},\ \bibinfo {pages} {504} (\bibinfo {year}
  {2022})}\BibitemShut {NoStop}%
\bibitem [{\citenamefont {Blundell}(2003)}]{blundell2003magnetism}%
  \BibitemOpen
  \bibfield  {author} {\bibinfo {author} {\bibfnamefont {S.}~\bibnamefont
  {Blundell}},\ }\href@noop {} {\enquote {\bibinfo {title} {Magnetism in
  condensed matter},}\ } (\bibinfo {year} {2003})\BibitemShut {NoStop}%
\bibitem [{\citenamefont {Overhauser}(1962)}]{overhauser1962spin}%
  \BibitemOpen
  \bibfield  {author} {\bibinfo {author} {\bibfnamefont {A.}~\bibnamefont
  {Overhauser}},\ }\href@noop {} {\bibfield  {journal} {\bibinfo  {journal}
  {Physical Review}\ }\textbf {\bibinfo {volume} {128}},\ \bibinfo {pages}
  {1437} (\bibinfo {year} {1962})}\BibitemShut {NoStop}%
\bibitem [{\citenamefont {Gr{\"u}ner}(1994)}]{gruner1994dynamics}%
  \BibitemOpen
  \bibfield  {author} {\bibinfo {author} {\bibfnamefont {G.}~\bibnamefont
  {Gr{\"u}ner}},\ }\href@noop {} {\bibfield  {journal} {\bibinfo  {journal}
  {Reviews of modern physics}\ }\textbf {\bibinfo {volume} {66}},\ \bibinfo
  {pages} {1} (\bibinfo {year} {1994})}\BibitemShut {NoStop}%
\bibitem [{\citenamefont {Lee}\ \emph {et~al.}(1974)\citenamefont {Lee},
  \citenamefont {Rice},\ and\ \citenamefont {Anderson}}]{lee1974conductivity}%
  \BibitemOpen
  \bibfield  {author} {\bibinfo {author} {\bibfnamefont {P.}~\bibnamefont
  {Lee}}, \bibinfo {author} {\bibfnamefont {T.}~\bibnamefont {Rice}}, \ and\
  \bibinfo {author} {\bibfnamefont {P.}~\bibnamefont {Anderson}},\ }\href@noop
  {} {\bibfield  {journal} {\bibinfo  {journal} {Solid State Communications}\
  }\textbf {\bibinfo {volume} {14}},\ \bibinfo {pages} {703} (\bibinfo {year}
  {1974})}\BibitemShut {NoStop}%
\bibitem [{\citenamefont {Mizuno}\ and\ \citenamefont
  {Izuyama}(1959)}]{mizuno1959electron}%
  \BibitemOpen
  \bibfield  {author} {\bibinfo {author} {\bibfnamefont {Y.}~\bibnamefont
  {Mizuno}}\ and\ \bibinfo {author} {\bibfnamefont {T.}~\bibnamefont
  {Izuyama}},\ }\href@noop {} {\bibfield  {journal} {\bibinfo  {journal}
  {Progress of Theoretical Physics}\ }\textbf {\bibinfo {volume} {21}},\
  \bibinfo {pages} {593} (\bibinfo {year} {1959})}\BibitemShut {NoStop}%
\bibitem [{\citenamefont {Huang}\ and\ \citenamefont
  {Maki}(1990)}]{huang1990imperfect}%
  \BibitemOpen
  \bibfield  {author} {\bibinfo {author} {\bibfnamefont {X.}~\bibnamefont
  {Huang}}\ and\ \bibinfo {author} {\bibfnamefont {K.}~\bibnamefont {Maki}},\
  }\href@noop {} {\bibfield  {journal} {\bibinfo  {journal} {Physical Review
  B}\ }\textbf {\bibinfo {volume} {42}},\ \bibinfo {pages} {6498} (\bibinfo
  {year} {1990})}\BibitemShut {NoStop}%
\bibitem [{\citenamefont {McMillan}(1975)}]{mcmillan1975landau}%
  \BibitemOpen
  \bibfield  {author} {\bibinfo {author} {\bibfnamefont {W.}~\bibnamefont
  {McMillan}},\ }\href@noop {} {\bibfield  {journal} {\bibinfo  {journal}
  {Physical Review B}\ }\textbf {\bibinfo {volume} {12}},\ \bibinfo {pages}
  {1187} (\bibinfo {year} {1975})}\BibitemShut {NoStop}%
\bibitem [{\citenamefont {Chen}\ \emph {et~al.}(2020)\citenamefont {Chen},
  \citenamefont {Ruan}, \citenamefont {Wu}, \citenamefont {Tang}, \citenamefont
  {Ryu}, \citenamefont {Tsai}, \citenamefont {Lee}, \citenamefont {Kahn},
  \citenamefont {Liou}, \citenamefont {Jia} \emph {et~al.}}]{chen2020strong}%
  \BibitemOpen
  \bibfield  {author} {\bibinfo {author} {\bibfnamefont {Y.}~\bibnamefont
  {Chen}}, \bibinfo {author} {\bibfnamefont {W.}~\bibnamefont {Ruan}}, \bibinfo
  {author} {\bibfnamefont {M.}~\bibnamefont {Wu}}, \bibinfo {author}
  {\bibfnamefont {S.}~\bibnamefont {Tang}}, \bibinfo {author} {\bibfnamefont
  {H.}~\bibnamefont {Ryu}}, \bibinfo {author} {\bibfnamefont {H.-Z.}\
  \bibnamefont {Tsai}}, \bibinfo {author} {\bibfnamefont {R.~L.}\ \bibnamefont
  {Lee}}, \bibinfo {author} {\bibfnamefont {S.}~\bibnamefont {Kahn}}, \bibinfo
  {author} {\bibfnamefont {F.}~\bibnamefont {Liou}}, \bibinfo {author}
  {\bibfnamefont {C.}~\bibnamefont {Jia}},  \emph {et~al.},\ }\href@noop {}
  {\bibfield  {journal} {\bibinfo  {journal} {Nature Physics}\ }\textbf
  {\bibinfo {volume} {16}},\ \bibinfo {pages} {218} (\bibinfo {year}
  {2020})}\BibitemShut {NoStop}%
\bibitem [{\citenamefont {Colonna}\ \emph {et~al.}(2005)\citenamefont
  {Colonna}, \citenamefont {Ronci}, \citenamefont {Cricenti}, \citenamefont
  {Perfetti}, \citenamefont {Berger},\ and\ \citenamefont
  {Grioni}}]{colonna2005mott}%
  \BibitemOpen
  \bibfield  {author} {\bibinfo {author} {\bibfnamefont {S.}~\bibnamefont
  {Colonna}}, \bibinfo {author} {\bibfnamefont {F.}~\bibnamefont {Ronci}},
  \bibinfo {author} {\bibfnamefont {A.}~\bibnamefont {Cricenti}}, \bibinfo
  {author} {\bibfnamefont {L.}~\bibnamefont {Perfetti}}, \bibinfo {author}
  {\bibfnamefont {H.}~\bibnamefont {Berger}}, \ and\ \bibinfo {author}
  {\bibfnamefont {M.}~\bibnamefont {Grioni}},\ }\href@noop {} {\bibfield
  {journal} {\bibinfo  {journal} {Physical review letters}\ }\textbf {\bibinfo
  {volume} {94}},\ \bibinfo {pages} {036405} (\bibinfo {year}
  {2005})}\BibitemShut {NoStop}%
\bibitem [{\citenamefont {Fazekas}(1999)}]{fazekas1999lecture}%
  \BibitemOpen
  \bibfield  {author} {\bibinfo {author} {\bibfnamefont {P.}~\bibnamefont
  {Fazekas}},\ }\href@noop {} {\emph {\bibinfo {title} {Lecture notes on
  electron correlation and magnetism}}},\ Vol.~\bibinfo {volume} {5}\ (\bibinfo
   {publisher} {World scientific},\ \bibinfo {year} {1999})\BibitemShut
  {NoStop}%
\bibitem [{\citenamefont {Kawamura}(2019)}]{kawamura2019fermisurfer}%
  \BibitemOpen
  \bibfield  {author} {\bibinfo {author} {\bibfnamefont {M.}~\bibnamefont
  {Kawamura}},\ }\href@noop {} {\bibfield  {journal} {\bibinfo  {journal}
  {Computer Physics Communications}\ }\textbf {\bibinfo {volume} {239}},\
  \bibinfo {pages} {197} (\bibinfo {year} {2019})}\BibitemShut {NoStop}%
\bibitem [{\citenamefont {Bao}\ \emph {et~al.}(2022)\citenamefont {Bao},
  \citenamefont {Cao}, \citenamefont {Krogstad}, \citenamefont {Taddei},
  \citenamefont {Shi}, \citenamefont {Cao}, \citenamefont {Lapidus},
  \citenamefont {Van~Smaalen}, \citenamefont {Chung}, \citenamefont
  {Kanatzidis} \emph {et~al.}}]{bao2022spin}%
  \BibitemOpen
  \bibfield  {author} {\bibinfo {author} {\bibfnamefont {J.-K.}\ \bibnamefont
  {Bao}}, \bibinfo {author} {\bibfnamefont {H.}~\bibnamefont {Cao}}, \bibinfo
  {author} {\bibfnamefont {M.~J.}\ \bibnamefont {Krogstad}}, \bibinfo {author}
  {\bibfnamefont {K.~M.}\ \bibnamefont {Taddei}}, \bibinfo {author}
  {\bibfnamefont {C.}~\bibnamefont {Shi}}, \bibinfo {author} {\bibfnamefont
  {S.}~\bibnamefont {Cao}}, \bibinfo {author} {\bibfnamefont {S.~H.}\
  \bibnamefont {Lapidus}}, \bibinfo {author} {\bibfnamefont {S.}~\bibnamefont
  {Van~Smaalen}}, \bibinfo {author} {\bibfnamefont {D.~Y.}\ \bibnamefont
  {Chung}}, \bibinfo {author} {\bibfnamefont {M.~G.}\ \bibnamefont
  {Kanatzidis}},  \emph {et~al.},\ }\href@noop {} {\bibfield  {journal}
  {\bibinfo  {journal} {Physical Review B}\ }\textbf {\bibinfo {volume}
  {106}},\ \bibinfo {pages} {L201111} (\bibinfo {year} {2022})}\BibitemShut
  {NoStop}%
\bibitem [{\citenamefont {Johannes}\ and\ \citenamefont
  {Mazin}(2008)}]{johannes2008fermi}%
  \BibitemOpen
  \bibfield  {author} {\bibinfo {author} {\bibfnamefont {M.}~\bibnamefont
  {Johannes}}\ and\ \bibinfo {author} {\bibfnamefont {I.}~\bibnamefont
  {Mazin}},\ }\href@noop {} {\bibfield  {journal} {\bibinfo  {journal}
  {Physical Review B}\ }\textbf {\bibinfo {volume} {77}},\ \bibinfo {pages}
  {165135} (\bibinfo {year} {2008})}\BibitemShut {NoStop}%
\bibitem [{\citenamefont {Johannes}\ \emph {et~al.}(2006)\citenamefont
  {Johannes}, \citenamefont {Mazin},\ and\ \citenamefont
  {Howells}}]{johannes2006fermi}%
  \BibitemOpen
  \bibfield  {author} {\bibinfo {author} {\bibfnamefont {M.}~\bibnamefont
  {Johannes}}, \bibinfo {author} {\bibfnamefont {I.}~\bibnamefont {Mazin}}, \
  and\ \bibinfo {author} {\bibfnamefont {C.}~\bibnamefont {Howells}},\
  }\href@noop {} {\bibfield  {journal} {\bibinfo  {journal} {Physical Review
  B}\ }\textbf {\bibinfo {volume} {73}},\ \bibinfo {pages} {205102} (\bibinfo
  {year} {2006})}\BibitemShut {NoStop}%
\bibitem [{\citenamefont {Whangbo}\ \emph {et~al.}(1991)\citenamefont
  {Whangbo}, \citenamefont {Canadell}, \citenamefont {Foury},\ and\
  \citenamefont {Pouget}}]{whangbo1991hidden}%
  \BibitemOpen
  \bibfield  {author} {\bibinfo {author} {\bibfnamefont {M.-H.}\ \bibnamefont
  {Whangbo}}, \bibinfo {author} {\bibfnamefont {E.}~\bibnamefont {Canadell}},
  \bibinfo {author} {\bibfnamefont {P.}~\bibnamefont {Foury}}, \ and\ \bibinfo
  {author} {\bibfnamefont {J.-P.}\ \bibnamefont {Pouget}},\ }\href@noop {}
  {\bibfield  {journal} {\bibinfo  {journal} {Science}\ }\textbf {\bibinfo
  {volume} {252}},\ \bibinfo {pages} {96} (\bibinfo {year} {1991})}\BibitemShut
  {NoStop}%
\bibitem [{\citenamefont {Moore}\ \emph {et~al.}(2010)\citenamefont {Moore},
  \citenamefont {Brouet}, \citenamefont {He}, \citenamefont {Lu}, \citenamefont
  {Ru}, \citenamefont {Chu}, \citenamefont {Fisher},\ and\ \citenamefont
  {Shen}}]{moore2010fermi}%
  \BibitemOpen
  \bibfield  {author} {\bibinfo {author} {\bibfnamefont {R.}~\bibnamefont
  {Moore}}, \bibinfo {author} {\bibfnamefont {V.}~\bibnamefont {Brouet}},
  \bibinfo {author} {\bibfnamefont {R.}~\bibnamefont {He}}, \bibinfo {author}
  {\bibfnamefont {D.}~\bibnamefont {Lu}}, \bibinfo {author} {\bibfnamefont
  {N.}~\bibnamefont {Ru}}, \bibinfo {author} {\bibfnamefont {J.-H.}\
  \bibnamefont {Chu}}, \bibinfo {author} {\bibfnamefont {I.}~\bibnamefont
  {Fisher}}, \ and\ \bibinfo {author} {\bibfnamefont {Z.-X.}\ \bibnamefont
  {Shen}},\ }\href@noop {} {\bibfield  {journal} {\bibinfo  {journal} {Physical
  Review B}\ }\textbf {\bibinfo {volume} {81}},\ \bibinfo {pages} {073102}
  (\bibinfo {year} {2010})}\BibitemShut {NoStop}%
\bibitem [{\citenamefont {Laverock}\ \emph {et~al.}(2005)\citenamefont
  {Laverock}, \citenamefont {Dugdale}, \citenamefont {Major}, \citenamefont
  {Alam}, \citenamefont {Ru}, \citenamefont {Fisher}, \citenamefont {Santi},\
  and\ \citenamefont {Bruno}}]{laverock2005fermi}%
  \BibitemOpen
  \bibfield  {author} {\bibinfo {author} {\bibfnamefont {J.}~\bibnamefont
  {Laverock}}, \bibinfo {author} {\bibfnamefont {S.}~\bibnamefont {Dugdale}},
  \bibinfo {author} {\bibfnamefont {Z.}~\bibnamefont {Major}}, \bibinfo
  {author} {\bibfnamefont {M.}~\bibnamefont {Alam}}, \bibinfo {author}
  {\bibfnamefont {N.}~\bibnamefont {Ru}}, \bibinfo {author} {\bibfnamefont
  {I.}~\bibnamefont {Fisher}}, \bibinfo {author} {\bibfnamefont
  {G.}~\bibnamefont {Santi}}, \ and\ \bibinfo {author} {\bibfnamefont
  {E.}~\bibnamefont {Bruno}},\ }\href@noop {} {\bibfield  {journal} {\bibinfo
  {journal} {Physical Review B}\ }\textbf {\bibinfo {volume} {71}},\ \bibinfo
  {pages} {085114} (\bibinfo {year} {2005})}\BibitemShut {NoStop}%
\bibitem [{\citenamefont {Knowles}\ \emph {et~al.}(2020)\citenamefont
  {Knowles}, \citenamefont {Yang}, \citenamefont {Muramatsu}, \citenamefont
  {Moulding}, \citenamefont {Buhot}, \citenamefont {Sayers}, \citenamefont
  {Da~Como},\ and\ \citenamefont {Friedemann}}]{knowles2020fermi}%
  \BibitemOpen
  \bibfield  {author} {\bibinfo {author} {\bibfnamefont {P.}~\bibnamefont
  {Knowles}}, \bibinfo {author} {\bibfnamefont {B.}~\bibnamefont {Yang}},
  \bibinfo {author} {\bibfnamefont {T.}~\bibnamefont {Muramatsu}}, \bibinfo
  {author} {\bibfnamefont {O.}~\bibnamefont {Moulding}}, \bibinfo {author}
  {\bibfnamefont {J.}~\bibnamefont {Buhot}}, \bibinfo {author} {\bibfnamefont
  {C.~J.}\ \bibnamefont {Sayers}}, \bibinfo {author} {\bibfnamefont
  {E.}~\bibnamefont {Da~Como}}, \ and\ \bibinfo {author} {\bibfnamefont
  {S.}~\bibnamefont {Friedemann}},\ }\href@noop {} {\bibfield  {journal}
  {\bibinfo  {journal} {Physical Review Letters}\ }\textbf {\bibinfo {volume}
  {124}},\ \bibinfo {pages} {167602} (\bibinfo {year} {2020})}\BibitemShut
  {NoStop}%
\bibitem [{\citenamefont {G{\"u}ller}\ \emph {et~al.}(2016)\citenamefont
  {G{\"u}ller}, \citenamefont {Vildosola},\ and\ \citenamefont
  {Llois}}]{guller2016spin}%
  \BibitemOpen
  \bibfield  {author} {\bibinfo {author} {\bibfnamefont {F.}~\bibnamefont
  {G{\"u}ller}}, \bibinfo {author} {\bibfnamefont {V.~L.}\ \bibnamefont
  {Vildosola}}, \ and\ \bibinfo {author} {\bibfnamefont {A.~M.}\ \bibnamefont
  {Llois}},\ }\href@noop {} {\bibfield  {journal} {\bibinfo  {journal}
  {Physical Review B}\ }\textbf {\bibinfo {volume} {93}},\ \bibinfo {pages}
  {094434} (\bibinfo {year} {2016})}\BibitemShut {NoStop}%
\bibitem [{\citenamefont {Sachdev}(2018)}]{sachdev2018topological}%
  \BibitemOpen
  \bibfield  {author} {\bibinfo {author} {\bibfnamefont {S.}~\bibnamefont
  {Sachdev}},\ }\href@noop {} {\bibfield  {journal} {\bibinfo  {journal}
  {Reports on Progress in Physics}\ }\textbf {\bibinfo {volume} {82}},\
  \bibinfo {pages} {014001} (\bibinfo {year} {2018})}\BibitemShut {NoStop}%
\bibitem [{\citenamefont {Dugdale}(2016)}]{dugdale2016life}%
  \BibitemOpen
  \bibfield  {author} {\bibinfo {author} {\bibfnamefont {S.~B.}\ \bibnamefont
  {Dugdale}},\ }\href@noop {} {\bibfield  {journal} {\bibinfo  {journal}
  {Physica Scripta}\ }\textbf {\bibinfo {volume} {91}},\ \bibinfo {pages}
  {053009} (\bibinfo {year} {2016})}\BibitemShut {NoStop}%
\bibitem [{\citenamefont {Abanov}\ and\ \citenamefont
  {Chubukov}(2000)}]{abanov2000spin}%
  \BibitemOpen
  \bibfield  {author} {\bibinfo {author} {\bibfnamefont {A.}~\bibnamefont
  {Abanov}}\ and\ \bibinfo {author} {\bibfnamefont {A.~V.}\ \bibnamefont
  {Chubukov}},\ }\href@noop {} {\bibfield  {journal} {\bibinfo  {journal}
  {Physical review letters}\ }\textbf {\bibinfo {volume} {84}},\ \bibinfo
  {pages} {5608} (\bibinfo {year} {2000})}\BibitemShut {NoStop}%
\bibitem [{\citenamefont {Bauer}\ \emph {et~al.}(2020)\citenamefont {Bauer},
  \citenamefont {Schattner}, \citenamefont {Trebst},\ and\ \citenamefont
  {Berg}}]{bauer2020hierarchy}%
  \BibitemOpen
  \bibfield  {author} {\bibinfo {author} {\bibfnamefont {C.}~\bibnamefont
  {Bauer}}, \bibinfo {author} {\bibfnamefont {Y.}~\bibnamefont {Schattner}},
  \bibinfo {author} {\bibfnamefont {S.}~\bibnamefont {Trebst}}, \ and\ \bibinfo
  {author} {\bibfnamefont {E.}~\bibnamefont {Berg}},\ }\href@noop {} {\bibfield
   {journal} {\bibinfo  {journal} {Physical review research}\ }\textbf
  {\bibinfo {volume} {2}},\ \bibinfo {pages} {023008} (\bibinfo {year}
  {2020})}\BibitemShut {NoStop}%
\bibitem [{\citenamefont {Eberlein}\ \emph {et~al.}(2016)\citenamefont
  {Eberlein}, \citenamefont {Metzner}, \citenamefont {Sachdev},\ and\
  \citenamefont {Yamase}}]{eberlein2016fermi}%
  \BibitemOpen
  \bibfield  {author} {\bibinfo {author} {\bibfnamefont {A.}~\bibnamefont
  {Eberlein}}, \bibinfo {author} {\bibfnamefont {W.}~\bibnamefont {Metzner}},
  \bibinfo {author} {\bibfnamefont {S.}~\bibnamefont {Sachdev}}, \ and\
  \bibinfo {author} {\bibfnamefont {H.}~\bibnamefont {Yamase}},\ }\href@noop {}
  {\bibfield  {journal} {\bibinfo  {journal} {Physical Review Letters}\
  }\textbf {\bibinfo {volume} {117}},\ \bibinfo {pages} {187001} (\bibinfo
  {year} {2016})}\BibitemShut {NoStop}%
\bibitem [{\citenamefont {Jacobsen}\ \emph {et~al.}(1982)\citenamefont
  {Jacobsen}, \citenamefont {Tanner},\ and\ \citenamefont
  {Bechgaard}}]{jacobsen1982optical}%
  \BibitemOpen
  \bibfield  {author} {\bibinfo {author} {\bibfnamefont {C.}~\bibnamefont
  {Jacobsen}}, \bibinfo {author} {\bibfnamefont {D.}~\bibnamefont {Tanner}}, \
  and\ \bibinfo {author} {\bibfnamefont {K.}~\bibnamefont {Bechgaard}},\
  }\href@noop {} {\bibfield  {journal} {\bibinfo  {journal} {Molecular Crystals
  and Liquid Crystals}\ }\textbf {\bibinfo {volume} {79}},\ \bibinfo {pages}
  {381} (\bibinfo {year} {1982})}\BibitemShut {NoStop}%
\end{thebibliography}%

%%%%%%%%%% Merge with supplemental materials %%%%%%%%%%

\widetext
\newpage
\begin{center}
\textbf{\large Supplemental Materials: Magnetic Stability, Fermi Surface Topology, and Spin-Correlated Dielectric Response in Monolayer 1T-CrTe$_2$}
\end{center}
%%%%%%%%%% Merge with supplemental materials %%%%%%%%%%
%%%%%%%%%% Prefix a "S" to all equations, figures, tables and reset the counter %%%%%%%%%%
\setcounter{equation}{0}
\setcounter{figure}{0}
\setcounter{table}{0}
\setcounter{subsection}{0}
\setcounter{page}{1}
\makeatletter
\renewcommand{\theequation}{S\arabic{equation}}
\renewcommand{\thefigure}{S\arabic{figure}}
\renewcommand{\thetable}{S\arabic{table}}
\renewcommand{\bibnumfmt}[1]{[S#1]}
\renewcommand{\citenumfont}[1]{S#1}
%%%%%%%%%% Prefix a "S" to all equations, figures, tables and reset the counter %%%%%%%%%%

\subsection{Energy Convergence With Respect To Sampling In The Unit Cell of Ferromagnetic CrTe$_2$}
Relaxation is performed on a unit cell until the force on each atom becomes less than 0.002 eV/$\angstrom$ and the total energy converges to within 10$^-8$ eV. A cutoff energy surface of 600 eV was used for the plane-wave-basis. A $\Gamma$-centered Monkhorst-Pack of $13\times13\times1$ was used to sample the Brillouin zone for relaxation. After relaxation, the mesh was changed to be $n\times n \times $1 where: $3 \leq n \leq 15$. Colinear self-consistent calculations were performed using different $n$ values to find a suitable mesh for the unit cell. 

In the figure below, the ground energy of the unit cell with respect to the size of a mesh $n \times n \times 1$ is shown. The value of $n$ was allowed to go up to 25. Notice how the energy oscillates around the value $E = -16.2625 $ eV, especially for values where $n \geq 10$. This oscillatory behavior is another indicator that the chosen mesh is suitable for the calculation. The mesh sizes of the supercells were picked with this criterion in mind. 
\begin{figure}[H]
    \centering
    \includegraphics[scale = 0.6]{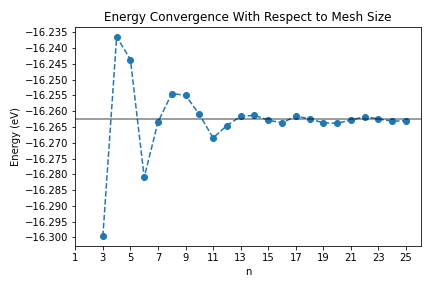}
    \caption{Energy Convergence as a function of sampling points.}
\end{figure}
%%%%%%%%%%%%%%------SECTION------%%%%%%%%%%%
\subsection{Ground Energy Convergence With Respect to Cutoff Energy In The Unit Cell of Ferromagnetic CrTe$_2$}
The relaxed structure generated in the previous section was also used to perform colinear self-consistent calculations to test the convergence of the ground state energy against the energy cutoff surface used for the plane-wave-basis set. A mesh of $13 \times 13\ \times 1$ was used for the calculation. The self-consistent calculations were performed until the total energy difference reached $10^{-8}$ eV.
\begin{table}[H]
\centering
\begin{tabular}{|c|c|c|}
	\hline
	E$_{cutoff} (eV) $ & E$_{total}$ (eV)\\
	\hline
	400 & -16.260964 \\
	\hline
	450 & -16.261668 \\
	\hline
	500 & -16.262638 \\
	\hline
	550 & -16.262849 \\
	\hline
	600 & -16.262793 \\
	\hline
	650 & -16.262816 \\
	\hline
	700 & -16.263001 \\
	\hline
\end{tabular} 
\caption{Ground Energy Convergence with respect to energy cutoff}
\end{table}

\subsection{Additional Supporting Figures}
The ground energy of the unit cell converges up to the 5$^{th}$ significant digit when with an energy cutoff surface of at least 500 eV. Hence, we conclude that calculations performed using $E_{cutoff} = 500$ eV or more are sufficient to model the behavior of the system.

\begin{figure*}[tbp]
\centering
\includegraphics[scale = 0.6]{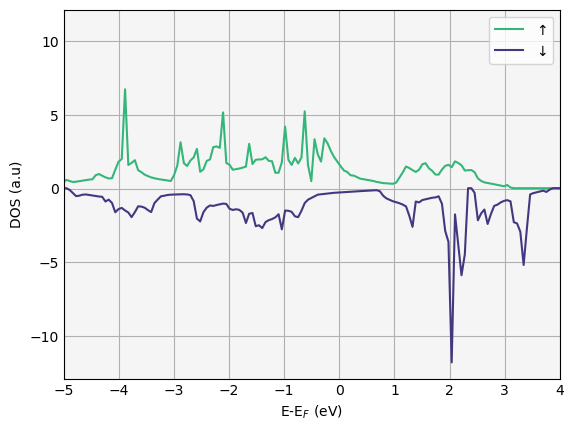}
\caption{DOS of the FM phase.}
\label{S1}
\end{figure*}

\begin{figure*}[tbp]
\centering
\includegraphics[scale = 0.6]{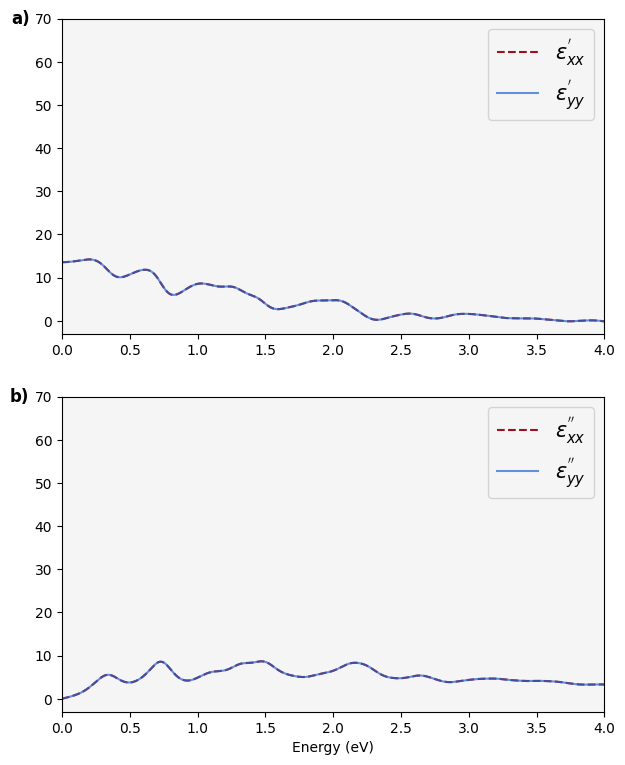}
\caption{Optical response of the FM unit cell along the $x$ and $y$ direction showing no anisotropy.}
\label{S2}
\end{figure*}

\begin{figure*}[tbp]
\centering
\includegraphics[scale = 0.6]{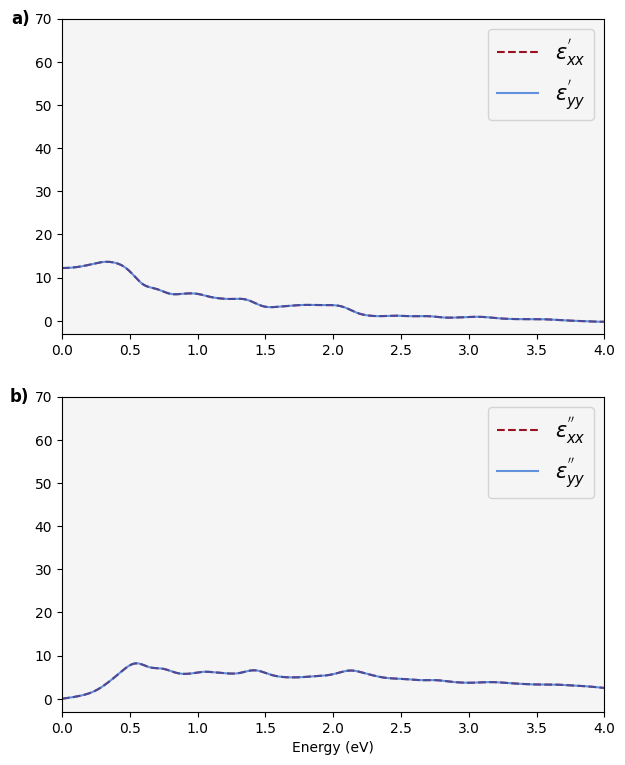}
\caption{Optical response of the FM-CDW supercell along the $x$ and $y$ direction showing no anisotropy.}
\label{S3}
\end{figure*}

\end{document}